\documentclass[12pt]{article}
\usepackage{amssymb,amsmath}
\usepackage{graphicx}
\usepackage{subfig}
\usepackage{color}
\usepackage{slashed}
\usepackage{enumerate}
\usepackage{dsfont}
\usepackage{hyperref}
\definecolor{nicered}{rgb}{0.7,0.1,0.1}
\definecolor{nicegreen}{rgb}{0.1,0.5,0.1}
\hypersetup{colorlinks,citecolor= nicegreen,linkcolor= nicered}
\allowdisplaybreaks
\begin{document}
\begin{titlepage}
\begin{flushright}
\end{flushright}
\newcommand{\AddrDOR}{{\sl \small Facult\"at f\"ur Physik, Technische 
    Universit\"at Dortmund\\ D-44221 Dortmund, Germany}}
\newcommand{\AddrLiege}{{\sl \small IFPA, Dep. AGO, 
    Universite de Liege, Bat B5,\\ \small \sl Sart
    Tilman B-4000 Liege 1, Belgium}}
\vspace*{1.5cm}
\begin{center}
  \textbf{\large The role of lepton flavor symmetries}\\[3mm]
  \textbf{\large in leptogenesis}
  \\[10mm]
  D. Aristizabal Sierra$^{a,}$\footnote{e-mail address: {\tt daristizabal@ulg.ac.be}},
  I. de Medeiros Varzielas$^{b,}$\footnote{e-mail address: {\tt ivo.de@udo.edu}}
  \vspace{0.8cm}
  \\
  $^a$\AddrLiege.\vspace{0.4cm}\\
  $^b$\AddrDOR.\vspace{0.4cm}  \\
\end{center}
\vspace*{0.5cm}
\begin{abstract}
  The presence of flavor symmetries in the lepton sector may have
  several consequences for the generation of the baryon asymmetry of
  the Universe via leptogenesis.  We review the mechanism in general
  type-I, type-II and type-III seesaw models. We then turn to the
  discussion of the cases when the asymmetry is generated in the
  context of seesaw models extended with flavor symmetries, before or
  after flavor symmetry breaking. Finally we explain how the interplay
  between type-I and type-II seesaws can (or not) lead to viable
  models for leptogenesis even when there is an exact mixing pattern
  enforced by the flavor symmetry.
\end{abstract}
\end{titlepage}
\setcounter{footnote}{0}
\section{Introduction}
\label{sec:intro}
Baryogenesis via leptogenesis is a scenario in which the baryon
asymmetry of the Universe is first generated in leptons and partially
reprocessed---via standard model sphaleron processes---into a baryon
asymmetry. From a general point of view three conditions (Sakharov
conditions \cite{Sakharov:1967dj}) have to be satisfied in order for
leptogenesis to take place at some stage during the evolution of the
expanding Universe, namely there must be ($i$) interactions that break
lepton number; ($ii$) CP violation; ($iii$) departure from
thermodynamical equilibrium. In principle any framework in which these
conditions can be satisfied can be regarded as a playground for
leptogenesis.

In models for Majorana neutrino masses lepton number is broken, so
they provide intrinsic frameworks for leptogenesis. The most well
studied scenarios for leptogenesis correspond to the type-I
\cite{Minkowski:1977sc,Yanagida:1979as,Glashow:1979nm,
  GellMann:1980vs,Mohapatra:1979ia,Schechter:1980gr}, type-II
\cite{Schechter:1980gr,Lazarides:1980nt,Mohapatra:1980yp,Wetterich:1981bx}
and type-III \cite{Foot:1988aq} seesaw models (tree-level realizations
of the dimension five effective operator ${\cal O}_5\sim \ell\ell HH$
\cite{Weinberg:1980bf}). In these cases the generation of a $B-L$
asymmetry proceeds via the decay of heavy fermion right-handed
electroweak singlets (RH neutrinos for brevity) (see
e.g. \cite{Davidson:2008bu}) or triplets \cite{Hambye:2003rt,Hambye:2005tk} (type-I
or type-III) or scalar $SU(2)$ triplets
\cite{Hambye:2003ka,Antusch:2004xy,Hambye:2005tk}. Due to the
different electroweak charges of these states their thermodynamical
behavior is different and so is the way in which leptogenesis takes
place.

The idea of flavor symmetries dates back to the late 1970's
\cite{Froggatt:1978nt}. Initially, due to the lack of experimental
data in the lepton sector, flavor symmetries were used to explain
quark masses and mixing patterns, but with the advent of neutrino data
\cite{Schwetz:2011zk,GonzalezGarcia:2010er,Fogli:2011qn} the idea was
increasingly extended to the lepton sector. In particular, in recent
years it has been shown that lepton mixing is well described by
non-Abelian flavor symmetries (see e.g. \cite{Altarelli:2010gt}). In
association with these developments, the issue of leptogenesis in
flavor models has attracted a great deal of attention.

In this short review we describe the relationship between flavor
symmetries and leptogenesis. In section \ref{sec:lepto-gen} we provide
a brief review of general aspects of leptogenesis, covering
leptogenesis in type-I seesaw, and also leptogenesis in type-II and
III seesaw models. We take some care in establishing the notation to
be used in the other sections. The connection with flavor symmetries
has been covered in several works:
\cite{AristizabalSierra:2007ur,AristizabalSierra:2009bh,Sierra:2011vk}
cover the flavor symmetric phase and are reviewed in section
\ref{sec:sym}. In section \ref{sec:single} we review the results of
\cite{Jenkins:2008rb,Hagedorn:2009jy,Bertuzzo:2009im,
  AristizabalSierra:2009ex,Felipe:2009rr,Choubey:2010vs}, addressing
the case where only type-I seesaw takes place and identify cases where
the presence of the symmetries can lead to strong predictions for the
viability of leptogenesis.  In section \ref{sec:both} we summarize the
results from \cite{AristizabalSierra:2011ab}, where the scenario
considered has both type-I and II seesaws taking place---here too, in
certain circumstances conclusions about leptogenesis can be derived
due to the presence of the flavor symmetry. Other papers studying
leptogenesis in the context of a flavor symmetry include
\cite{Mohapatra:2005ra,Antusch:2006cw,Adhikary:2008au,Lin:2009ic,Branco:2009by,Altarelli:2009kr,Riva:2010jm}
although here we do not review their results explicitly.

For consistency we employ the same notation throughout the review. The
notation is mostly based on what was used in
\cite{AristizabalSierra:2009ex} and parts of the notation used in
\cite{AristizabalSierra:2011ab} (in particular where type-II seesaw is
discussed, such as in section \ref{sec:both}). In general we consider
the basis where the charged lepton mass matrix is diagonal. For type-I seesaw we also take the basis where the RH neutrino
mass matrix is diagonal unless otherwise stated. Matrices that appear with a hat are in the
basis where that matrix is diagonal (e.g. $\pmb{\hat{m}_D}$) and we
denote matrices in boldface.

\section{Leptogenesis: generalities}
\label{sec:lepto-gen}
We now discuss the general framework of leptogenesis in more detail:
in a hot plasma with $N$ lepton number and CP violating states
$S_1,\cdots, S_N$, assuming they all have tree-level couplings with
the standard model leptons (only with electroweak doublets to simplify
the discussion), their out-of-equilibrium tree-level decays will
produce a net $B-L$ asymmetry. The determination of the exact amount
of $B-L$ asymmetry depends on the dynamics of the $S_\alpha$ states
and requires---in general---an analysis based on kinetic equations
accounting for the evolution of the $S_\alpha$ densities and the $B-L$
asymmetry density itself. For the evolution of the $B-L$ asymmetry one
can write
\begin{equation}
  \label{eq:total-lepton-asymm}
  \dot Y_{\Delta_{B-L}}(z)=\sum_{\alpha=1}^N\;
  \dot Y_{\Delta_{B-L}}^{(S_\alpha)}(z)\,,
\end{equation}
where, following ref. \cite{Nardi:2007jp}, we are using the notation
$s(z)H(z)z \,d\,Y_X(z)/dz\equiv \dot Y_X(z)$. Here $z=M_1/T$ ($M_1$
being the mass of the lightest state), $Y_{\Delta_X}=n_X-n_{\bar X}/s$
with $n_X$ ($n_{\bar X}$) the number density of particles
(antiparticles), $s$ the entropy density and $H(z)$ the expansion rate
of the Universe (the expressions for these functions are given in
appendix \ref{sec:conventions}). $\dot Y_{\Delta_{B-L}}^{(S_\alpha)}(z)$ is
the asymmetry generated by each of the states $S_\alpha$. Note that we
have written the dimensionless inverse temperature of the remaining
states as $z_\alpha=M_\alpha/M_1\, z$.

The evolution of the asymmetries generated by each $S_\alpha$ is in
turn determined by the ``competition'' between source ($S_{S_\alpha}$)
and washout ($W_{S_\alpha}$) terms. The size of the source terms is
fixed by how much the $S_\alpha$'s deviate from thermodynamical
equilibrium when decaying, by the strength of the decays and by the
amount of CP violation. The size of the washout terms is, instead,
dictated by $S_\alpha$ processes that tend to diminish the lepton
asymmetry created via the source terms like e.g. inverse decays and
lepton number breaking scatterings.

As discussed in the introduction, models for Majorana neutrino masses are
intrinsic scenarios for leptogenesis to take place and indeed from this approach it turns out that there is
a link between two in principle unrelated problems: the origin of
neutrino masses and the origin of the baryon asymmetry of the
Universe. It is well known that Majorana neutrino masses can be
generated in a model independent way by adding to the standard model
Lagrangian an effective dimension five operator ${\cal O}_5\sim\ell
\ell H H$, that generates the corresponding Majorana masses after
electroweak symmetry breaking \cite{Weinberg:1980bf}. The tree-level
realizations of this operator give rise to type-I, II and III seesaws
which constitute the usual frameworks for almost all the studies of
leptogenesis.
\subsection{Leptogenesis in type-I seesaw}
\label{sec:leptogenesis-type-I}
In type-I seesaw the states $S_\alpha$ correspond to RH
neutrinos $N_{\alpha}$. In a general basis the interactions of these
states are given by
\begin{equation}
  \label{eq:seesaw-lag}
  -{\cal L}^{(I)}=
  \text{i}\bar N\,\gamma^\mu\partial_\mu\,N
  + \bar \ell\,\pmb{\lambda}^*\,N \tilde H
  + \frac{1}{2} N^T\,C\,\pmb{M_N}\,N + \mbox{h.c.}\,,
\end{equation}
Here $\tilde H = \text{i}\tau_2 H^*$, $C$ is the charge conjugation
operator, and for 3 $N_{\alpha}$, $\pmb{\lambda}^*$ is a $3\times 3$ Yukawa coupling matrix in
flavor space and $\pmb{M_N}$ is the $3\times 3$ Majorana mass matrix.  At
energy scales well below the RH neutrino masses, the light neutrinos
masses are determined by the effective matrix
\begin{equation}
  \label{eq:eff-mass-matrix-type-I}
  \pmb{m_\nu^\text{eff}}=\pmb{m_\nu^I}
  =-\pmb{m_D}\pmb{M_{N}}^{-1}\pmb{m_D}^T
  =-\sum_{\alpha=1,2,3}M_{N_\alpha}^{-1}
  \pmb{m_{D_\alpha}}\otimes\pmb{m_{D_\alpha}}\,,
\end{equation}
where in order to facilitate the discussion in section \ref{sec:both}
we have expressed the matrix in terms of the parameter space vectors
$\pmb{m_{D_\alpha}}^T=(m_{D_{\alpha 1}},m_{D_{\alpha 2}},m_{D_{\alpha
    3}})$, with $\pmb{m_D}=v\,\pmb{\lambda}$ and $v=\langle
H\rangle\simeq 174$ GeV.  Diagonalization of
(\ref{eq:eff-mass-matrix-type-I}) by means of the PMNS mixing matrix
$\pmb{U}$ leads to the light neutrino mass spectrum:
\begin{equation}
  \label{eq:diagonalization-type-I}
  \pmb{U}^T\pmb{m_\nu^\text{eff}}\pmb{U}=\pmb{\hat m_\nu}\,,
\end{equation}
with $\pmb{U}=\pmb{V}\,\pmb{D}$ (with $\pmb{V}$ the part of the
PMNS matrix having a CKM-like form and $\pmb{D}=\mbox{diag}(e^{\text{i}\phi_1},e^{\text{i}\phi_2},1)$ containing
the Majorana CP phases).

The $3\times 3$ Dirac mass matrix $\pmb{m_D}$, being a general complex
matrix, contains 18 parameters (9 moduli and 9 phases) of which 3
phases can be removed by rotation of the lepton doublets in
(\ref{eq:seesaw-lag}). The number of physical parameters defining
$\pmb{m_D}$ is therefore 15. A very useful parametrization in which
this is explicitly taken into account is the Casas-Ibarra
parametrization \cite{Casas:2001sr}, in which the Dirac mass matrix is
expressed in terms of low-energy neutrino observables and a general
complex orthogonal matrix $\pmb{R}$, namely
\begin{equation}
  \label{eq:casas-ibarra}
  \pmb{m_D}=\pmb{U}^*\,\pmb{\hat m_\nu}^{1/2}\pmb{R}\,\pmb{\hat M_N}^{1/2}\,.
\end{equation}

\begin{figure}
  \centering
  \includegraphics[width=9cm,height=2cm]{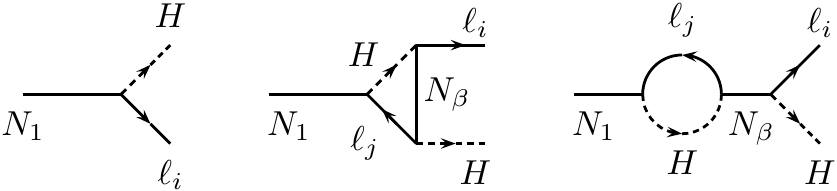}
  \caption{Feynman diagrams generating the CP violating asymmetry 
    $\epsilon_{N_1}^{\ell_i}$ in type-I seesaw.}
  \label{fig:cp-asymm-type-I}
\end{figure}

In the conventional thermal leptogenesis scenario the RH neutrino mass
spectrum is taken to be hierarchical, $M_{N_\alpha}\ll M_{N_\beta}$ for
$\alpha<\beta$ (for a throughout review see
\cite{Davidson:2008bu}). Under this simplification---well justified as
far as $T_\text{Reheat}<M_{N_{2,3}}$---the effects of $N_{2,3}$ can be
neglected and thus the asymmetry is entirely produced by the dynamics
of $N_1$. The kinetic equations that describe the evolution of the
asymmetry involve $N_1$ decays, $\Delta L=1$ and $\Delta L=2$
scatterings, and depending on the temperature regimen at which
leptogenesis takes place ($T\sim M_{N_1}$) should include the lepton
flavor degrees of freedom
\cite{Barbieri:1999ma,Endoh:2003mz,Fujihara:2005pv,Nardi:2006fx,
Abada:2006fw,Abada:2006ea}.
At the leading order in the Yukawa couplings, however, the kinetic
equations are determined by the decays and the off-shell pieces of the
$\Delta L=2$ scattering processes $\ell_j H\leftrightarrow \ell_i H$
and $\bar\ell_j H^\dagger\leftrightarrow \ell_i H$. For $T\gtrsim
10^{12}$ GeV (or otherwise neglecting flavor effects) they read as
follows
\begin{align}
  \label{eq:BEQ-type-I}
  \dot Y_{N_1}&=-(y_{N_1}-1)\gamma_{D_{N_1}}\,,\nonumber\\
  \dot Y_{\Delta_{B-L}}&=S_{N_1} + W_{N_1}\nonumber\\
  &=-\left[
    (y_{N_1}-1)\epsilon_{N_1}
    + \frac{y_{\Delta_{B-L}}}{2}
  \right]\gamma_{D_{N_1}}\,,
\end{align}
where we are using the notation $y_X\equiv Y_X/Y_X^\text{Eq}$ and
$y_{\Delta_{B-L}}\equiv Y_{\Delta_{B-L}}/Y_\ell^\text{Eq}$ (the
expressions for the equilibrium densities are given in appendix
\ref{sec:conventions}). The strength of the decays
\begin{equation}
  \label{eq:mtilde-type-I}
  \tilde m_1=\frac{v^2}{M_{N_1}}\,
  \left(
    \pmb{\lambda}^\dagger\pmb{\lambda}
  \right)_{11}
\end{equation}
determines the size of the reaction density $\gamma_{D_{N_1}}$
appearing in the source term $S_{N_1}$ as well as in the washout term
$W_{N_1}$, namely
\begin{equation}
  \label{eq:reaction-dens-type-I}
  \gamma_{D_{N_1}}=\frac{1}{8\pi^3}\frac{M_{N_1}^5}{v^2}
  \frac{K_1(z)}{z}\,\tilde m_1\,.
\end{equation}
Here $K_1(z)$ is the first-order modified Bessel function of the
second-type. The amount of CP violation in $N_1$ decays is given by
the CP violating asymmetry
$\epsilon_{N_1}=\sum_{i=e,\mu,\tau}\epsilon_{N_1}^{\ell_i}$. At the
leading order this quantity arises through the interference between
the $N_1$ tree-level decay and the one-loop vertex and wave function
corrections \cite{Covi:1996wh}. The flavored CP violating asymmetries
arising from the diagrams depicted in figure \ref{fig:cp-asymm-type-I}
read
\begin{align}
  \label{eq:CPV-asymmetries-type-I}
  \epsilon_{N_1}^{\ell_i\text{(V)}}&=\frac{1}{8\pi}\sum_{\beta\neq 1}
  \frac{\mathbb{I}\mbox{m}[\sqrt{\omega_\beta}
    (\pmb{\lambda}^\dagger\pmb{\lambda})_{\beta 1}
    \lambda_{i\beta}^*\lambda_{i1}]}
  {(\pmb{\lambda}^\dagger\pmb{\lambda})_{11}}f(\omega_\beta)
  \;,\nonumber\\
  \epsilon_{N_1}^{\ell_i\text{(W)}}&=
  -\frac{1}{8\pi}\sum_{\beta\neq 1}
  \frac{
    \mathbb{I}\mbox{m}
    \{
    [(\pmb{\lambda}^\dagger\pmb{\lambda})_{1\beta}
    +
    \sqrt{\omega_\beta}(\pmb{\lambda}^\dagger\pmb{\lambda})_{\beta 1}]
    \lambda_{i\beta}^*\lambda_{i1}
    \}}
  {(\pmb{\lambda}^\dagger\pmb{\lambda})_{11}}g(\omega_\beta)\;,
\end{align}
with obvious generalization if the decaying state is $N_\alpha$, where
$\omega_\beta=M_{N_\beta}^2/M_{N_\alpha}^2$ and the loop functions are
\begin{align}
  \label{eq:loop-functions-type-I}
  f(\omega_\beta)&=(1+\omega_\beta)\ln
  \left(
    \frac{\omega_\beta + 1}{\omega_\beta}
  \right) - 1\;,\nonumber\\
  g(\omega_\beta)&=\frac{1}{\omega_\beta-1}\;.
\end{align}
Since $\omega_\beta^{-1}\ll 1$ the loop functions can be expanded in
powers of $\omega_\beta^{-1}$ dropping the subleading terms. At
leading order in these parameters the flavored CP violating
asymmetries can be expressed as
\begin{equation}
  \label{eq:flavored-cpv-asymmetry-leading-type-I}
  \epsilon_{N_1}^{\ell_i}=
  -\frac{1}{8\pi(\pmb{\lambda}^\dagger\pmb{\lambda})_{11}}
  \sum_{\beta\neq 1}
  \mathbb{I}\mbox{m}
  \left\{
    \left[
      \frac{(\pmb{\lambda}^\dagger\pmb{\lambda})_{1\beta}}{\omega_\beta}
      + 
      \frac{3(\pmb{\lambda}^\dagger\pmb{\lambda})_{\beta 1}}{2\sqrt{\omega_\beta}}
    \right]
    \lambda_{i\beta}^*\lambda_{i 1}
    \right\}\,.
\end{equation}
The total CP asymmetry then is obtained from
\begin{equation}
  \label{eq:total-CP-type-I}
  \epsilon_{N_1}=\sum_{i=e,\mu,\tau}\epsilon_{N_1}^{\ell_i}=
  -\frac{3}{16\pi v^2}\sum_\beta\frac{1}{\sqrt{\omega_\beta}}
  \frac{\mathbb{I}\mbox{m}[(\pmb{m_D}^\dagger\pmb{m_D})_{\beta 1}^2]}
  {(\pmb{m_D}^\dagger\pmb{m_D})_{11}}\;,
\end{equation}
where anticipating the discussions of sections \ref{sec:single} and
\ref{sec:both} we have rewritten the CP violating asymmetry in terms
of the Dirac mass matrix (in the basis where the RH neutrino mass
matrix is diagonal).

\begin{figure}
  \centering
  \includegraphics[width=9cm,height=7cm]{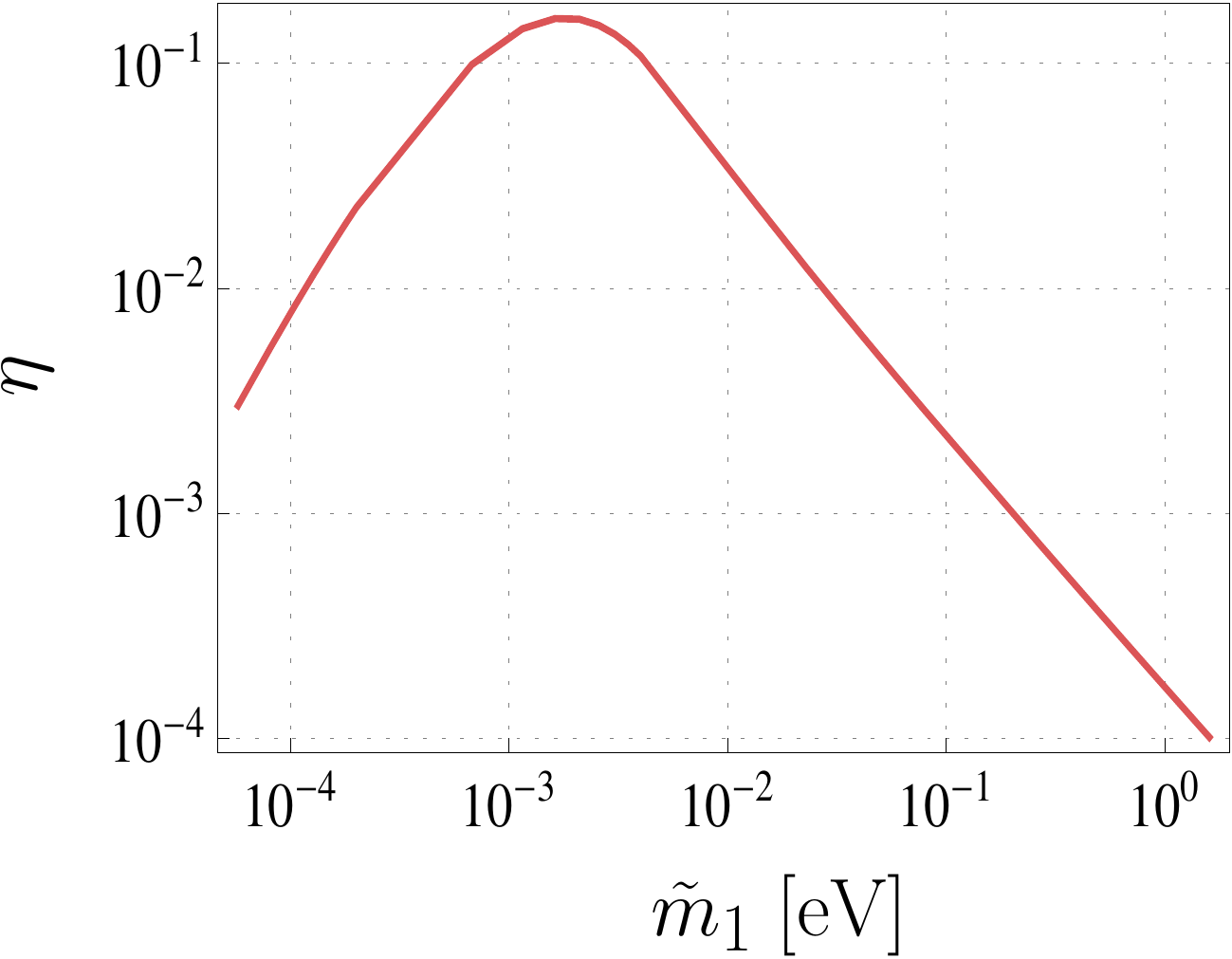}
  \caption{Efficiency factor as a function of the parameter $\tilde
    m_1$ in type-I seesaw.  The efficiency has been calculated from
    leading order Boltzmann equations in the one-flavor
    approximation.}
  \label{fig:efficiency-type-I}
\end{figure}

Formal integration of eqs. (\ref{eq:BEQ-type-I}), using the
integrating factor technique and assuming a vanishing primordial $B-L$
asymmetry, gives
\begin{equation}
  \label{eq:BmL-astmm-type-I}
  Y_{\Delta_{B-L}}(z)=-\epsilon_{N_1}\,Y_{N_1}^\text{Eq}(z\to 0)\,\eta(z)\,.
\end{equation}
Here $\eta(z)$ is the efficiency function that determines the
evolution of $Y_{\Delta_{B-L}}(z)$ and its final value at $z\to\infty$
(see appendix \ref{sec:conventions} for details). At ${\cal
  O}(\pmb{\lambda}^2)$ (leading order) the problem of determining the
final $Y_{\Delta_{B-L}}$ is a two parameters problem, $\tilde m_1$ and
$\epsilon_{N_1}$, and requires numerical solutions of
eqs. (\ref{eq:BEQ-type-I}) \footnote{Analytically the problem has been
  addressed yielding quite accurate expressions for the efficiency
  \cite{Buchmuller:2004nz,Blanchet:2006dq}.}. Figure
\ref{fig:efficiency-type-I} shows the efficiency factor
$\eta\equiv\eta(z\to \infty)$ as a function of the parameter $\tilde
m_1$.
\subsection{Leptogenesis in type-II seesaw}
\label{sec:leptogenesis-type-II}
Consistent models for leptogenesis involving scalar electroweak
triplets require going beyond a single scalar triplet.  There are
several ways in which this can be done, namely adding at least another
triplet \cite{Ma:1998dx}, adding RH neutrinos \cite{Hambye:2003ka} or
adding fermionic triplets. Here we will discuss scenarios that include
RH neutrinos, as these schemes will be further analyzed in the context
of flavor symmetries in section \ref{sec:both}. In this case,
depending on the triplet and RH mass spectrum, the states $S_\alpha$
can be identified with the RH neutrinos, the triplet or both.

The interactions of the RH neutrinos are given by the Lagrangian in
(\ref{eq:seesaw-lag}) whereas the interaction of the triplet are determined
by
\begin{equation}
  \label{eq:Lag-type-II}
  -{\cal L}^{(II)}=\ell^T C \pmb{Y}\text{i}\tau_2\pmb{\tau}\ell
  \pmb{\Delta}
  +M^2_\Delta \mbox{Tr}\pmb{\Delta}^\dagger\pmb{\Delta}
  -\mu H^T \text{i}\tau_2 \pmb{\tau}H \pmb{\Delta}+ \mbox{h.c.}\,.
\end{equation}
Here $\pmb{Y}$ is a $3\times 3$ matrix in flavor space and
$\pmb{\Delta}$, the $SU(2)$ scalar electroweak triplet has hypercharge
+1 (to the lepton doublets -1/2) and is given by
\begin{equation}
  \label{eq:triplet}
  \pmb{\Delta}=
  \begin{pmatrix}
    \Delta^{++} & \Delta^{+}/\sqrt{2}\\
    \Delta^{+}/\sqrt{2} & \Delta^{0}
  \end{pmatrix}\,.
\end{equation}
After electroweak symmetry breaking the light neutrino mass matrix
receives the contributions from the dimension five effective operators of the RH neutrino and
triplet
\begin{equation}
  \label{eq:eff-mass-matrix}
  \pmb{m_\nu^{\text{eff}}}=\pmb{m_\nu^I} + \pmb{m_\nu^{II}}\quad\text{with}\quad
  \pmb{m_\nu^{II}}=2\,v_\Delta\,\pmb{Y}\,.
\end{equation}
The first term is the contribution from the RH neutrinos
(eq.~(\ref{eq:eff-mass-matrix-type-I})) whereas the second one is the
contribution from the triplet, with the triplet vacuum expectation
value fixed by
$\langle\Delta^0\rangle=v_\Delta=\mu^*\,v^2/M_\Delta^2$.

As already mentioned the generation of $B-L$ asymmetry depends on the
heavy mass spectrum. One can define three possible scenarios:
\begin{itemize}
\item $M_{N_1}\ll M_\Delta$: the effects of $\pmb{\Delta}$ are
decoupled and the lepton asymmetry is generated via $N_1$
dynamics. This case resembles leptogenesis in type-I seesaw.
\item $M_{N_1}\gg M_\Delta$: the lepton asymmetry is entirely produced
  by the dynamics of $\pmb{\Delta}$ \cite{Hambye:2005tk}.
\item $M_\Delta\sim M_{N_1}$: both the triplet and the lightest RH
  neutrino generate the asymmetry \cite{AristizabalSierra:2011ab}.
\end{itemize}
Here we will discuss the third possibility in the regimen
$M_{N_1,\Delta}>10^{12}$ GeV.  We will closely follow the presentation
in \cite{AristizabalSierra:2011ab}.

Since in this case the scalar triplet, carrying non-trivial $SU(2)$
quantum numbers, couples to the standard model electroweak gauge
bosons, and the number of degrees of freedom participating in the
generation of the lepton asymmetry is larger, the Boltzmann equations
are more involved. At leading order in the Yukawa couplings
$\pmb{\lambda}$ and $\pmb{Y}$ the kinetic equation for the lepton
asymmetry, $Y_{\Delta_L}$\footnote{In contrast to the previous
  section, in this case we do not include the change in the lepton
  densities due to sphaleron processes, and thus study only the
  evolution of the $L$ asymmetry.}, involve the RH neutrino and
triplet decays and inverse decays $N_1\leftrightarrow \ell \tilde
H^\dagger$ and $\pmb{\Delta}\leftrightarrow \bar \ell\bar \ell$ and
the off-shell Yukawa generated scattering reactions $\ell \tilde
H^\dagger\leftrightarrow \ell \tilde H^\dagger$ and $H^\dagger
H^\dagger \leftrightarrow \ell\ell$. In addition to the evolution of
the $Y_{\Delta_L}$ asymmetry the full network of Boltzmann equations
should include the equations accounting for the evolution of the RH
neutrino and triplet number densities and the triplet and Higgs
asymmetries\footnote{These asymmetries are a consequence of these
  fields not being self-conjugate.}. The resulting system of five
coupled differential equations can be reduced to four by using the
constraint imposed by hypercharge neutrality \cite{Hambye:2005tk}:
\begin{equation}
  \label{eq:hypercharge-neutrality}
  2\,Y_{\Delta_\Delta} + Y_{\Delta H} - Y_{\Delta_L}=0\,.
\end{equation}

The resulting kinetic equations can thus be written as
\begin{align}
  \label{eq:BEQ-type-II}
  \dot Y_{N_1}&=-(y_{N_1}-1)\,\gamma_{D_{N_1}}\,,
  \nonumber\\
  \dot Y_\Sigma&=-(y_\Sigma - 1)\,\gamma_{D_\Delta} 
  - 2(y_\Sigma^2 - 1)\,\gamma_A\,,\nonumber\\
  \dot Y_{\Delta_L}&=\left[(y_{N_1}-1)\,\epsilon_{N_1}^\text{tot} -
    \left(y_{\Delta_L} + y_{\Delta_\Delta}^H\right)\right]\,\gamma_{D_{N_1}}
  +\left[(y_\Sigma - 1)\,\epsilon_\Delta -
    2K_\ell\, (y_{\Delta_L} + y_{\Delta_\Delta})\right]\,\gamma_{D_\Delta}\,,
  \nonumber\\
  \dot Y_{\Delta_\Delta}&=-\left[y_{\Delta_\Delta} + (K_\ell - K_H)\,y_{\Delta_L}
  + 2K_H \,y_{\Delta_\Delta}^H\right]\,,
\end{align}
where $\Sigma\equiv\Delta + \Delta^\dagger$ and $y_{\Delta_{\Delta}}^H\equiv
Y_{\Delta_\Delta}/Y_H^\text{Eq}$ and the rest of the variables in the
equations follow the conventions introduced in the previous section
when writing the eqs. in (\ref{eq:BEQ-type-I}). The reaction densities
involving the triplet are given by
\begin{equation}
  \label{eq:reaction-densities-triplet}
  \gamma_{D_\Delta}=\frac{1}{8\,\pi^3}\frac{M_\Delta^5}{v^2}
  \frac{K_1(z)}{z}\,
  \left(
    \tilde m_\Delta^\ell
    +
    \frac{\tilde m_\Delta^2}{4\tilde m_\Delta^\ell}
  \right)\,,
  \quad
  \gamma_A(z)=\frac{M_\Delta^4}{64\,\pi^4}\,\int_4^\infty
  dx\sqrt{x}\frac{K_1(zx)}{z}\,\widehat\sigma_A(x)\,,
\end{equation}
with $x=s/M_\Delta^2$. The reduced cross section
$\widehat\sigma_A(x)=2\,x\,\lambda(1,x^{-1},0)$ (where we have
$\lambda(a,b,c)=(a-b-c)^2-4bc$) can be found in appendix
\ref{sec:conventions}.  
The factors $K_{\ell,H}$ resemble the flavor projectors
defined in standard flavored leptogenesis
\cite{Nardi:2006fx,Abada:2006ea} as they project triplet decays
into either the Higgs or the lepton doublet directions. They are
defined as follows
\begin{equation}
  \label{eq:projectors}
  K_\ell=\frac{\tilde m_\Delta^\ell}{\tilde m_\Delta^\ell + \frac{\tilde m_\Delta^2}
    {4\,\tilde m_\Delta^\ell}}\,
  \qquad\mbox{and}\qquad
  K_H=\frac{\tilde m_\Delta^2}{4\,\tilde m_\Delta^\ell\left(
    \tilde m_\Delta^\ell + \frac{\tilde m_\Delta^2}
    {4\,\tilde m_\Delta^\ell}\right)}\,,
\end{equation}
where the parameters $\tilde m_\Delta^\ell$ and $\tilde m^2_\Delta$
are given by
\begin{equation}
  \label{eq:definition-mtildes}
  \tilde m_\Delta^\ell=\frac{v^2\,|\pmb{Y}|^2}{M_\Delta}
  \qquad\mbox{and}\qquad
  \tilde m^2_\Delta=\mbox{Tr}[\pmb{m_\nu^{II}}\pmb{m_\nu^{II}}^\dagger]\,,
\end{equation}
with $|\pmb{Y}|^2=\mbox{Tr}[\pmb{Y}\,\pmb{Y}^\dagger]$.  In these
definitions we have replaced the trilinear coupling $\mu$ by the
contribution of the type-II sector to the effective light neutrino
mass matrix, encoded in $\tilde m^2_\Delta$. In principle this is just
a matter of choice, but it proves to be quite convenient given that in
contrast to $\mu$ the parameter $\tilde m_\Delta$ is (partially)
constrained by experimental neutrino data.

\begin{figure}
  \centering
  \includegraphics[width=7cm,height=2.4cm]{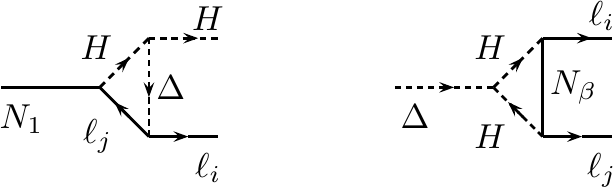}
  \caption{Left-hand side: Vertex loop correction involving the
    triplet and contributing to $\epsilon_{N_1}^\Delta$. Right-hand
    side: Vertex loop correction for triplet decays
    \cite{Hambye:2003ka}.}
  \label{fig:cp-asymm-type-II}
\end{figure}

The CP asymmetry for the RH neutrino arises as in type-I but, due to
the trilinear scalar coupling in (\ref{eq:Lag-type-II}), there is an
additional contribution coming from a vertex correction involving the
triplet, as shown in fig. \ref{fig:cp-asymm-type-II} (left-hand
side). The interference between the tree-level decay $N_\alpha\to
\ell\,\tilde H^\dagger$ and this 1-loop vertex diagram yields
\cite{Hambye:2003ka,Antusch:2004xy}\footnote{This equation follows
  from \cite{Antusch:2004xy} which differs from \cite{Hambye:2003ka}
  by a factor of $3/2$.}
\begin{equation}
  \label{eq:cp-asymm-triplet}
    \epsilon_{N_1}^\Delta=-\frac{3}{2\,\pi\,M_\Delta}
  \frac{1}{
    \left(
      \pmb{m_D}\,\pmb{m_D}^\dagger
    \right)_{11}}
  \mathfrak{I}\mbox{m}
  \left[
    \left(\pmb{m_D}\,\pmb{Y}^*\pmb{m_D}^T
    \right)_{11}\,\mu
  \right]\,h(\sigma_1)\,.
\end{equation}
The function $h(\sigma_1)$, with $\sigma_\alpha=M_\Delta^2/M_{N_\alpha}^2$, is
given by
\begin{equation}
  \label{eq:triplet-cp-asymm-loop-function}
  h(\sigma_1)=\sqrt{\sigma_1}
  \left[
    1 - \sigma_1\log
    \left(\frac{1+\sigma_1}{\sigma_1}\right)
  \right]\,.
\end{equation}
The total CP violating asymmetry in $N_1$ decays therefore reads
\begin{equation}
  \label{eq:total-CP-N1-type-II}
  \epsilon_{N_1}^\text{tot}=\epsilon_{N_1}+\epsilon_{N_1}^\Delta\,,
\end{equation}
where, for the scenario considered, $\epsilon_{N_1}$ is determined by
eqs.~(\ref{eq:flavored-cpv-asymmetry-leading-type-I}) and
(\ref{eq:total-CP-type-I}).

The CP violating asymmetry in triplet decays arises from
the interference between the tree-level $\pmb{\Delta}\to \ell\ell$
process and the interference with the 1-loop vertex diagram shown in
figure \ref{fig:cp-asymm-type-II} (right-hand side). The result reads
\cite{Hambye:2003ka}
\begin{equation}
  \label{eq:triplet-CPasymmetry}
  \epsilon_\Delta=-\frac{1}{8\,\pi\,v^2}\frac{1}{M_\Delta}
  \frac{
    \sum_\beta\mathfrak{I}\mbox{m}
    \left[
      \left(
        \pmb{m_D}\,\pmb{Y}^*\,\pmb{m_D}^T
      \right)_{\beta\beta}\,\mu
    \right]}{\mbox{Tr}\left[\pmb{Y}\,\pmb{Y}^\dagger\right] + \mu^2/M_\Delta^2}\,
  H(\sigma_\beta)\,,
\end{equation}
where the loop function in this case is given by
\begin{equation}
  \label{eq:loop-function-triplet-decay}
  H(\sigma_\beta)=\frac{1}{\sqrt{\sigma_\beta}}
  \log\left(1 + \sigma_\beta \right)\,.
\end{equation}

As in the type-I case the kinetic equation for the lepton asymmetry
can be formally integrated. The resulting asymmetry, assuming a zero
primordial asymmetry, can be expressed in two different ways
\cite{AristizabalSierra:2011ab}
\begin{equation}
  \label{eq:efficiencies}
  Y_{\Delta_L}(z)=-\epsilon_{N_1}^\text{tot}\,Y^\text{Eq}_\text{tot}\,\eta^I(z)
  \quad\mbox{or}\quad
  Y_{\Delta_L}(z)=-\epsilon_\Delta\,Y^\text{Eq}_\text{tot}\,\eta^{II}(z)\,.
\end{equation}
The functions $\eta^{I,II}(z)$ are defined in such a way that in the
limit in which the triplet (RH neutrino) interactions are absent
$\eta^{I}$ ($\eta^{II}$) corresponds to the efficiency function of
standard leptogenesis (pure triplet leptogenesis), see appendix
\ref{sec:conventions} for details. As in the type-I case the final $L$
asymmetry is obtained from these functions in the limit $z\to \infty$.

A precise determination of the lepton asymmetry generated in $N_1$ and
$\pmb{\Delta}$ decays requires solving the network of equations in
(\ref{eq:BEQ-type-II}). Taking $z=M_\Delta/T$ and $z_N=r z$, with
$r=\sigma_1^{-1/2} = M_{N_1}/M_\Delta$, and once the CP asymmetries
$\epsilon_{N_1}^\text{tot}$ and $\epsilon_\Delta$ are fixed, the
problem of studying the evolution of the lepton asymmetry is entirely
determined by five parameters: $\tilde m_1$, $\tilde m_\Delta$, 
$\tilde m_\Delta^\ell$, $M_\Delta$ and $r$ \footnote{This
  is to be compared with the pure triplet leptogenesis scenario
  \cite{Hambye:2005tk} where the generation of the $L$ asymmetry is
  entirely determined by only three parameters: $\tilde m_\Delta$,
  $\tilde m_\Delta^\ell$, $M_\Delta$.}.

\begin{figure}
  \centering
  \includegraphics[width=7.4cm,height=6cm]{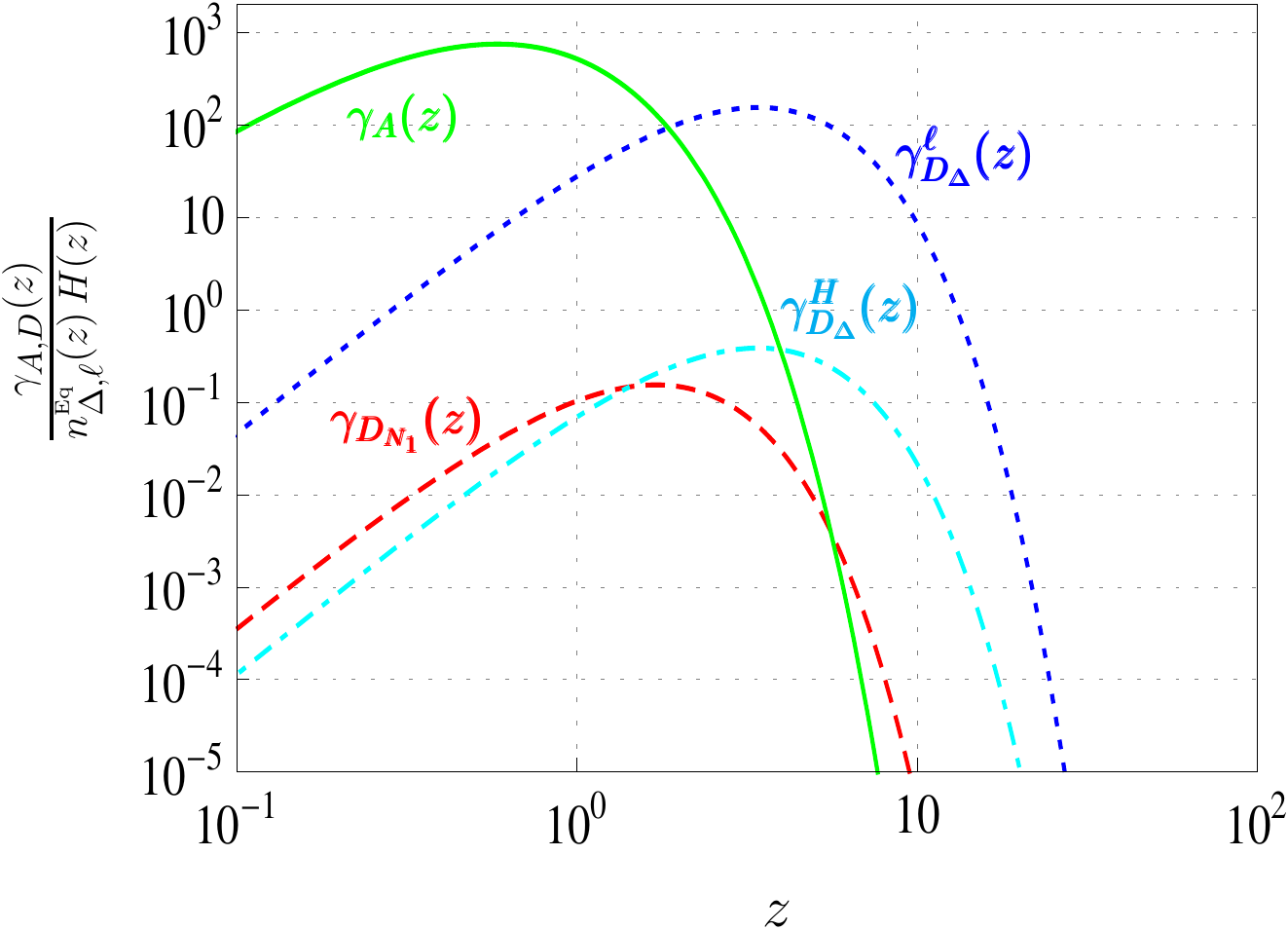}
  \includegraphics[width=7.4cm,height=6cm]{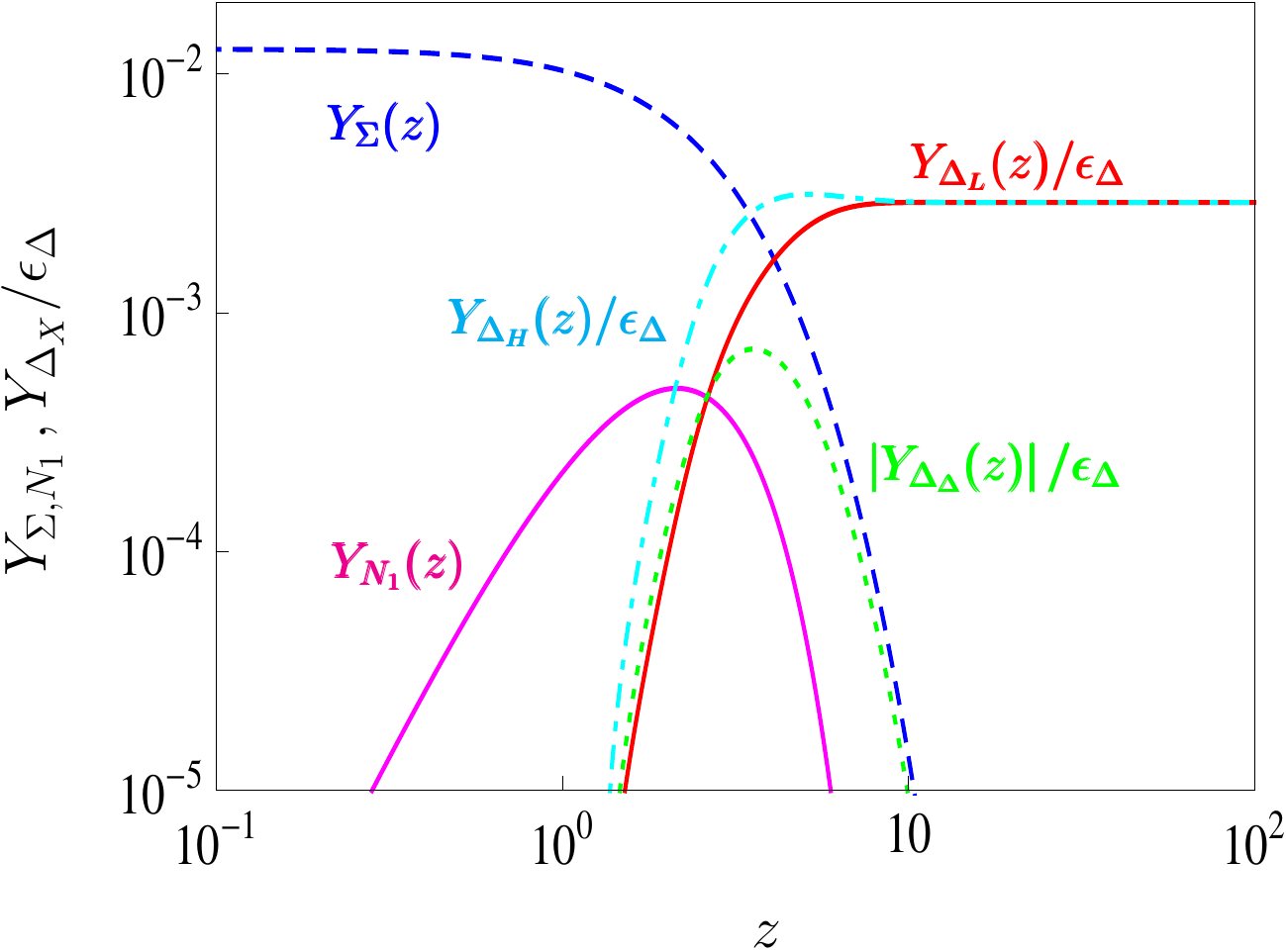}
      \includegraphics[width=7.4cm,height=6cm]{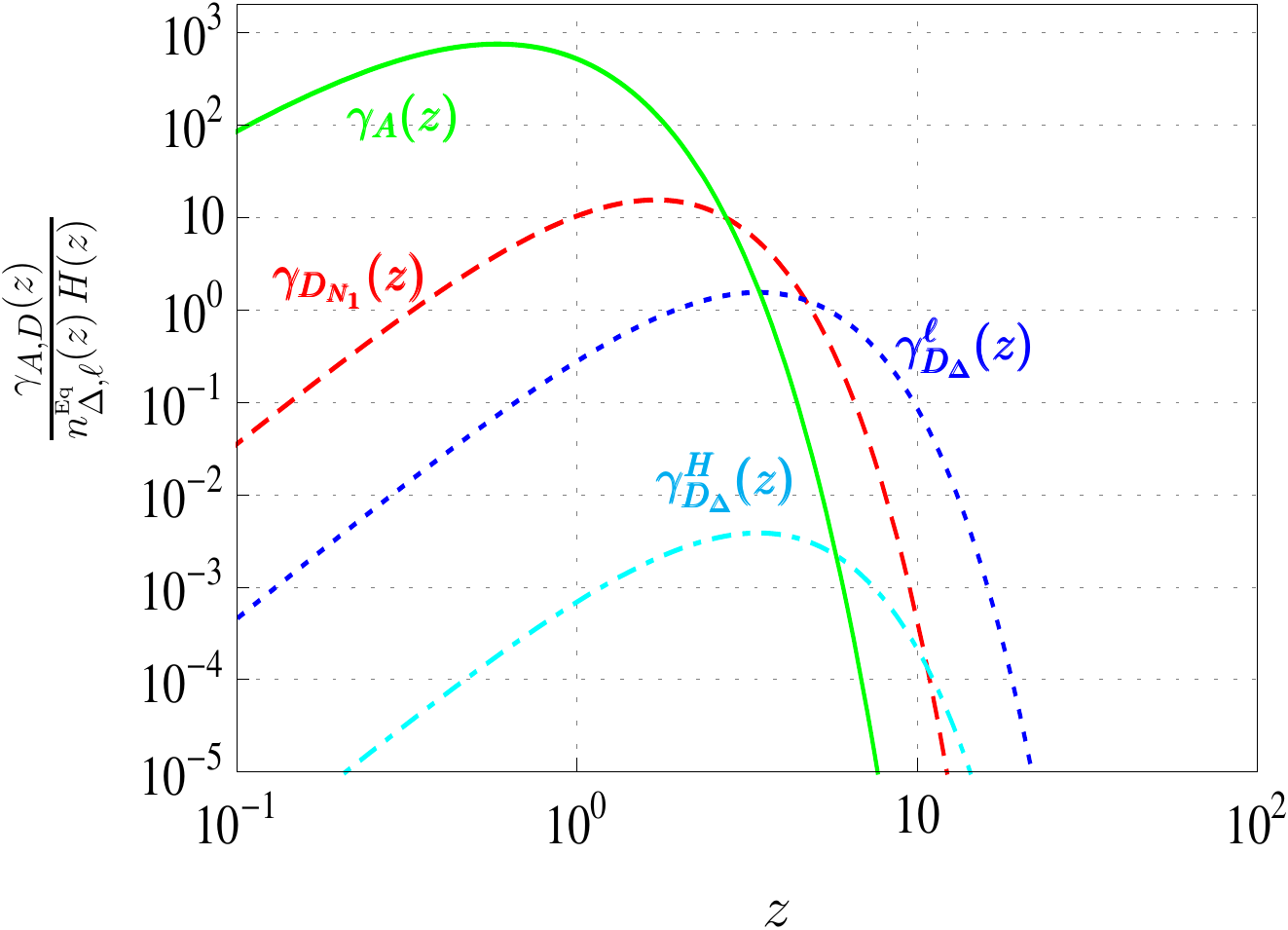}
    \includegraphics[width=7.4cm,height=6cm]{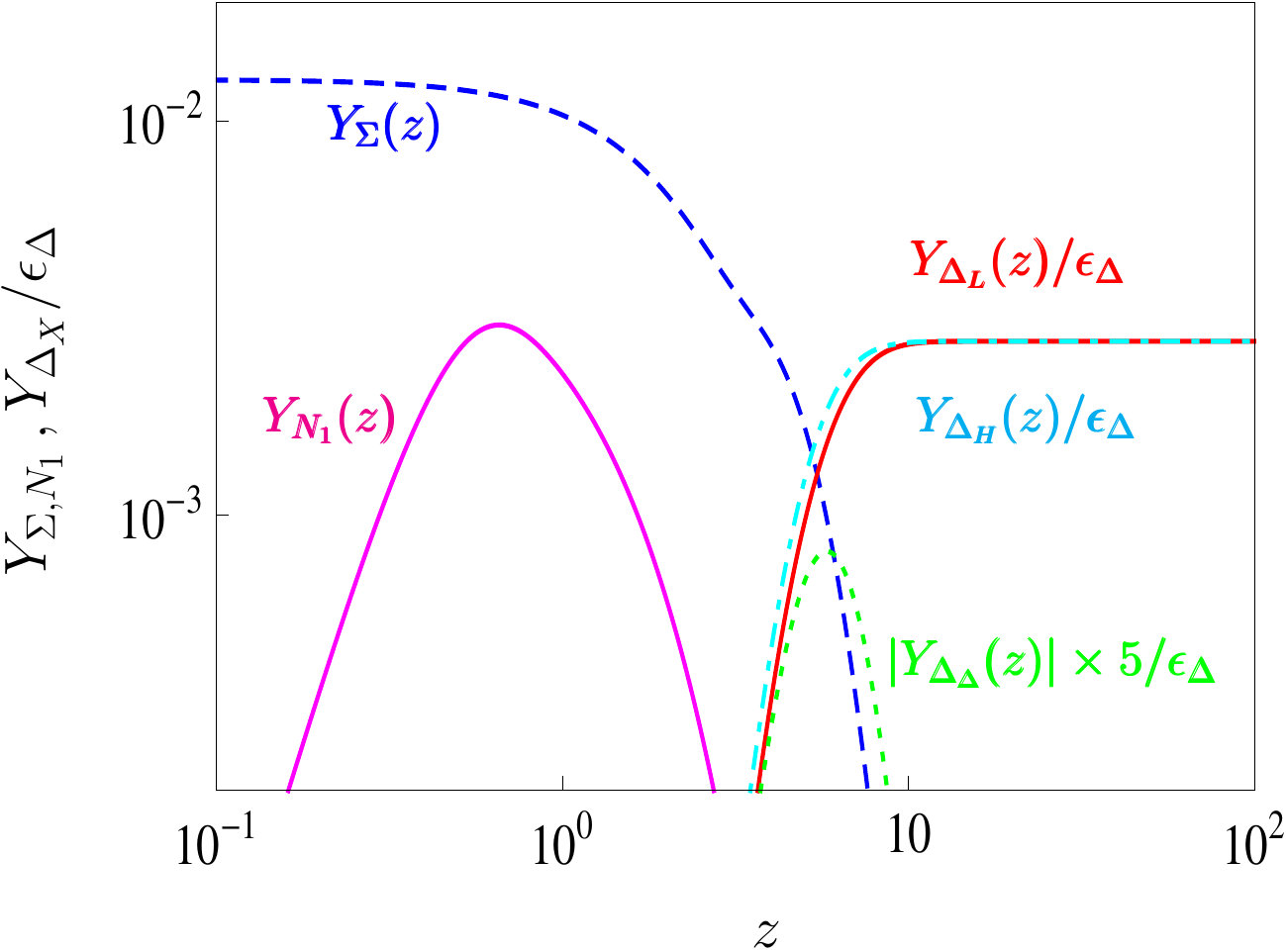}
    \caption{{\it Reaction densities for triplet and RH neutrino
        processes (left-panels) and evolution of the different
        densities (right-panels) entering in the kinetic equations for
        the scenarios of purely triplet leptogenesis (upper panels)
        and singlet dominated models (lower panels)
        \cite{AristizabalSierra:2011ab}. See the text for more
        details.}}
  \label{fig:models-case1-case2}
\end{figure}

As pointed out in \cite{AristizabalSierra:2011ab}, in models
featuring a mild hierarchy between $M_\Delta$ and $M_{N_1}$ three
scenarios can be defined:
\begin{enumerate}[I.]
\item \underline{Purely triplet scalar leptogenesis models:}\\
  The relevant parameters follow the hierarchy $\tilde m_1\ll
  \tilde m_\Delta^\ell, \tilde m_\Delta$. The $L$ asymmetry is
  generated through the processes $\pmb{\Delta}\to \bar\ell\bar\ell$
  or $\pmb{\Delta}\to H H$ and the details strongly depend on whether
  $\tilde m_\Delta^\ell\gg \tilde m_\Delta$, $\tilde m_\Delta^\ell\ll
  \tilde m_\Delta$ or $\tilde m_\Delta^\ell\sim \tilde m_\Delta$.
  Interestingly, when $\tilde m_\Delta^\ell\gg \tilde m_\Delta$ the
  Higgs asymmetry---being weakly washed out---turns out to be large
  and implies a large lepton asymmetry.
\item \underline{Singlet dominated leptogenesis models:}\\
  These scenarios are defined according to $\tilde m_1\gg \tilde
  m_\Delta^\ell, \tilde m_\Delta$ thus leptogenesis is mainly
  determined by $N_1$ dynamics. The relative difference between the
  parameters $\tilde m_\Delta^\ell$ and $\tilde m_\Delta$ determines
  whether either the Higgs asymmetry or the $L$ asymmetry are strongly
  or weakly washed out, thus three cases can be distinguished: $\tilde
  m_\Delta^\ell\gg \tilde m_\Delta$, $\tilde m_\Delta^\ell\ll \tilde
  m_\Delta$ or $\tilde m_\Delta^\ell\sim \tilde m_\Delta$. Each of
  them exhibit different features.
\item \underline{Mixed leptogenesis models:}\\
  In these models the parameters controlling the gauge reaction
  densities strengths are all of the same order i.e.  $\tilde
  m_1\sim \tilde m_\Delta^\ell\sim \tilde m_\Delta$.
\end{enumerate}
For the sake of illustration in figure \ref{fig:models-case1-case2} we
show two numerical examples for scenarios I and II. They were obtained
with the parameter space points $P_I$=($\tilde m_1$, $\tilde
m_\Delta$, $\tilde m_\Delta^\ell$,$M_\Delta$,$r$) =($10^{-4}$ eV,
$10^{-2}$ eV, $10^{-1}$ eV, $10^{10}$ GeV,2) and $P_{II}$=($\tilde
m_1$, $\tilde m_\Delta$, $\tilde m_\Delta^\ell$,$M_\Delta$,$r$)
=($10^{-2}$ eV, $10^{-4}$ eV, $10^{-3}$ eV, $10^{10}$ GeV,2) for fixed
$\epsilon_\Delta=10^{-6}$ and $\epsilon_{N_1}=10^{-5}$ and assuming
initial vanishing asymmetries.

\subsection{Leptogenesis in type-III seesaw}
\label{sec:leptogenesis-type-III}
In type-III \cite{Foot:1988aq} seesaw the states $S_\alpha$ correspond
to fermion electroweak triplets (here we consider 3 for definiteness) with vanishing
hypercharge. In a general basis the interactions of these states are
given by the following Lagrangian
\begin{equation}
  \label{eq:lagrangian-type-III}
  -{\cal L}^{(III)}=-\overline{\pmb{T}}_\alpha \slashed{D}\,\pmb{T}_\alpha
  + \overline{\ell}\,\pmb{h}^*\pmb{T} \widetilde H
  + \frac{1}{2}
  \overline{\pmb{T}}_\alpha^\dagger C \pmb{M_{T}} \pmb{T}_\alpha
  + \mbox{h.c.}\;,
\end{equation}
where the fermion triplets can be written as a matrix
\begin{equation}
  \label{eq:electric-triplet}
  \pmb{T}_\alpha = \pmb{\tau} \cdot \pmb{T}_\alpha=
  \begin{pmatrix}
    T_\alpha^0         & \sqrt{2}T_\alpha^+\\
    \sqrt{2}T_\alpha^- & -T_\alpha^0
  \end{pmatrix}\,,
\end{equation}
with $T^0 = T^3$, $T^\pm =(T^1\mp \text{i} T^2)/\sqrt{2}$. In this notation,
the covariant derivative is defined as $D_\mu = \partial_\mu - \text{i} g
\tau^a W_\mu^a / 2$ ($a$ being $SU(2)$ indices). Lepton number is
broken by the Majorana triplet mass terms and the effective light
neutrino mass matrix has the same structure than in type-I seesaw,
eq. (\ref{eq:eff-mass-matrix-type-I}), with the right-handed neutrino
mass matrix replaced by that of the triplets and $\pmb{m_D}=v\,\pmb{h}$:
\begin{equation}
  \label{eq:light-neutrino-mm-typeIII}
  \pmb{m_\nu^\text{eff}}=\pmb{m_\nu^{III}}=
  \sum_{\alpha=1,2,3}M_{T_\alpha}^{-1}
  \pmb{m_{D_\alpha}}\otimes\pmb{m_{D_\alpha}}\,,
\end{equation}
where we are using the same conventions used in the type-I case
discussion.

In what concerns leptogenesis, in several aspects, these models
resemble models based on type-I seesaw. For example assuming a
hierarchical triplet mass spectrum $M_{T_\alpha}<M_{T_\beta}$
($\alpha<\beta$) the $B-L$ asymmetry is completely produced by $T_1$
decays. There is, however, a significant difference arising from the fact
that the triplets couple to the standard model electroweak gauge
bosons. Thus, at high temperatures the triplet distribution is
thermalized by gauge reactions, and only when these reactions are
frozen a net $B-L$ asymmetry can be built
\cite{Hambye:2003rt,Strumia:2008cf}. 

As done in sections \ref{sec:leptogenesis-type-I} and
\ref{sec:leptogenesis-type-II}, in what follows, we will discuss the
generation of the $B-L$ asymmetry in these models in the one-flavor
approximation (the effects of flavor have been considered in
\cite{AristizabalSierra:2010mv}). At ${\cal O}(\pmb{h}^2)$, the
leading order in the couplings $\pmb{h}$, the kinetic equations
consist of $T_1$ decays and off-shell $\Delta L=2$ processes. The main
difference with the conventional leptogenesis scenario is the
inclusion of the couplings of $T_1$ with gauge bosons. The Boltzmann
equations in this case read
\begin{align}
  \label{eq:BEQs-type-III}
  \dot Y_{T_1}&=-\left(y_{T_1}-1\right)\gamma_{D_{T_1}}
  -\left(y_{T_1}^2-1\right)\gamma_A\,,\nonumber\\
  \dot Y_{\Delta_{B-L}}&=-
  \left[
    \left(y_{T_1}-1\right)\epsilon_{T_1}
    +\frac{y_{\Delta_{B-L}}}{2}
  \right]\gamma_{D_{T_1}}\,.
\end{align}
The Yukawa reaction density $\gamma_{D_{T_1}}$ is given by
eq. (\ref{eq:reaction-dens-type-I}), changing $\pmb{\lambda}\to
\pmb{h}$ and $M_{N_1}\to M_{T_1}$ in the definition of $\tilde m_1$
(eq. (\ref{eq:mtilde-type-I})) whereas the gauge reaction density by
(\ref{eq:reaction-densities-triplet}) using, of course, the
corresponding fermion triplet reduced cross section (see appendix
\ref{sec:conventions}). The CP violating asymmetry is a factor of
three smaller than in type-I seesaw due to contractions of the $SU(2)$
indices in the Yukawa interaction terms entering in the 1-loop
corrections, thus
\begin{equation}
  \label{eq:CP-asym-type-III}
  \epsilon_{T_1}=\sum_{i=e,\mu,\tau}\epsilon_{T_1}^{\ell_i}=
  -\frac{1}{16\pi v^2}\sum_\beta\frac{1}{\sqrt{\omega_\beta}}
  \frac{\mathbb{I}\mbox{m}[(\pmb{m_D}^\dagger\pmb{m_D})_{\beta 1}^2]}
  {(\pmb{m_D}^\dagger\pmb{m_D})_{11}}\;.
\end{equation}

From the formal integration of the $B-L$ asymmetry kinetic equation in
(\ref{eq:BEQs-type-III}) the asymmetry can be written as
\begin{equation}
  \label{eq:BmL-astmm-type-III}
  Y_{\Delta_{B-L}}(z)=-3 \epsilon_{T_1}\,Y_{T_1}^\text{Eq}(z\to 0)
  \,\eta(z)\,.
\end{equation}
The expression is similar to the one obtained in the type-I
case but the efficiency is different, as it now includes the gauge
reaction density. The factor of $3$ comes from the $SU(2)$ degrees
of freedom of $T_1$.

\begin{figure}
  \centering
  \includegraphics[height=6.4cm,width=7.4cm]{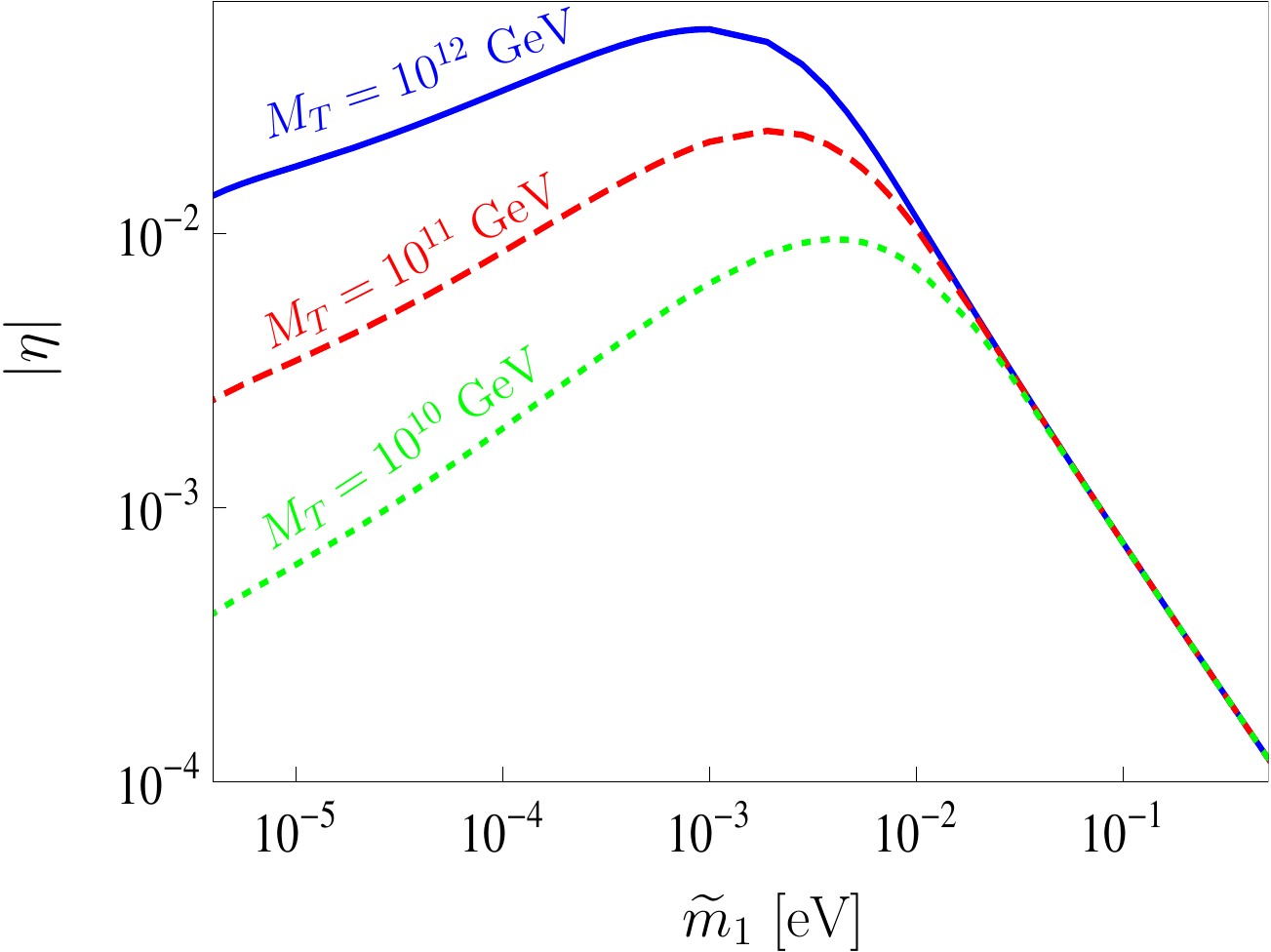}
  \includegraphics[height=6.4cm,width=7.4cm]{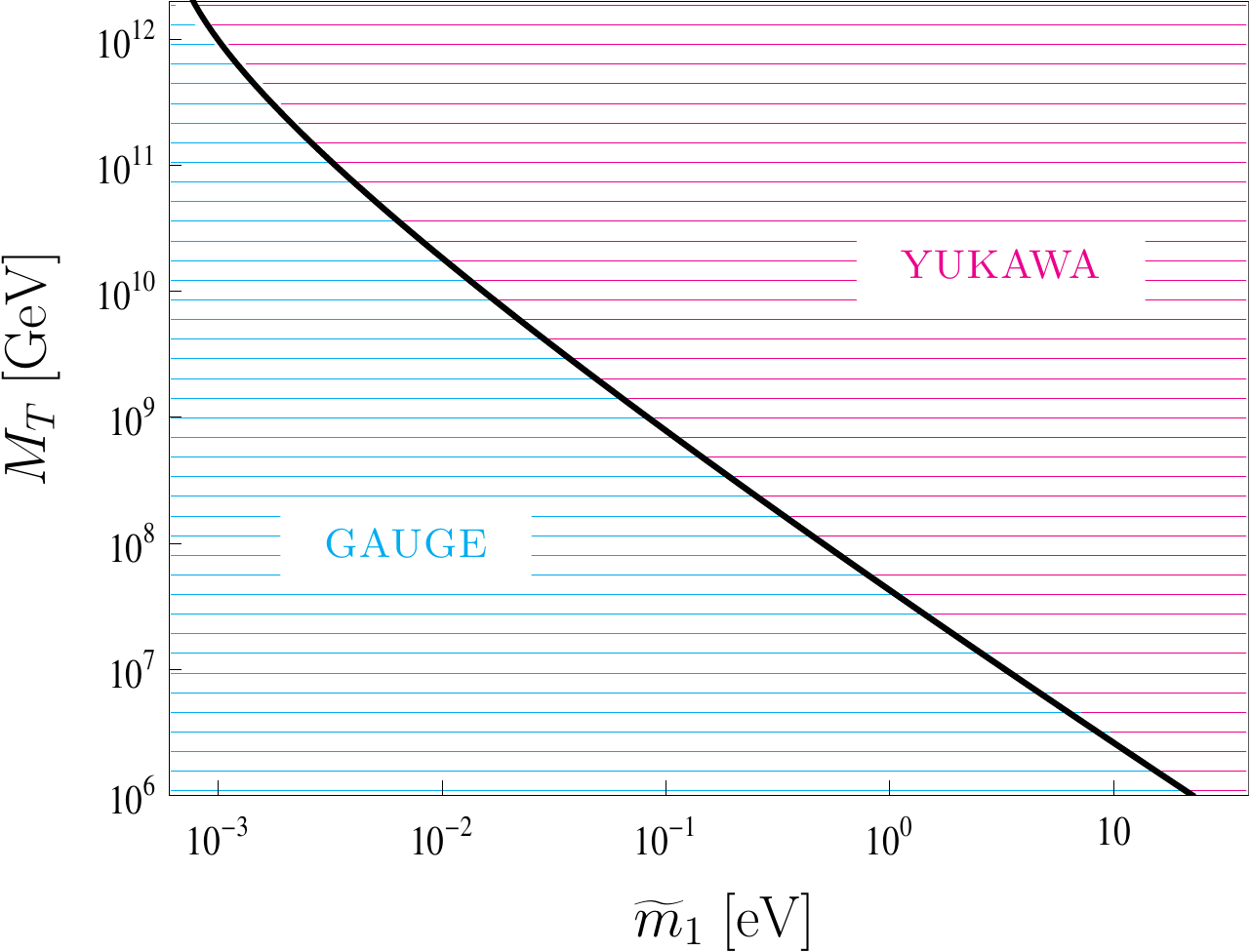}
  \caption{Efficiency factor as a function of $\widetilde m_1$ in the
    flavor aligned case (left panel) and regions for which gauge
    interactions freeze out after (lower region) and before (upper
    region) Yukawa reaction decoupling (right panel) in type-III
    leptogenesis.}
  \label{fig:eff-mtilde-aligned-type-III}
\end{figure}

A precise determination of the $B-L$ asymmetry relies on numerical
solutions of the kinetic equations, which in this case---even at the
leading order in the couplings---requires
$\epsilon_{T_1}$, $\tilde m_1$ and also the triplet mass
$M_{T_1}$ to be specified. The results for the efficiency factor are shown in
fig.~\ref{fig:eff-mtilde-aligned-type-III} (left panel) where a strong
dependence with $M_{T_1}$ can be seen. This dependence, introduced by
the gauge reactions, diminishes as $\tilde m_1$ increases and
disappears at certain $\widetilde m_1^{\text{min}}$. This implies that
above this value $T_1$ leptogenesis proceeds as in type-I seesaw
\footnote{In standard leptogenesis at ${\cal O}(\pmb{\lambda}^2)$ the
  efficiency does not depend on the RH neutrino mass.}. Thus, as
highlighted in \cite{AristizabalSierra:2010mv}, in this
type of models the generation of the $B-L$ asymmetry can proceed
either in in a region determined by the condition $\widetilde
m_1<\widetilde m_1^{\text{min}}$ (``gauge region'') or conversely in a
region defined by $\widetilde m_1>\widetilde m_1^{\text{min}}$
(``Yukawa region''). These regions are displayed in figure
\ref{fig:eff-mtilde-aligned-type-III} (right panel).
%

\section{Leptogenesis in the flavor symmetric phase}
\label{sec:sym}
We now turn to the discussion of the implications of the presence of
lepton flavor symmetries for leptogenesis in models based on type-I
seesaw. In general in these models four energy
scales can be distinguished: a cutoff scale $\Lambda$ (typically
a scale of heavy matter), the lepton number breaking scale $M_N$,
the flavons scale $M_\phi$---determined by the scale of the fields
that trigger flavor symmetry breaking---and the scale at which the
flavor symmetry is broken, denoted hereafter by $v_F$.  The scale of
heavy matter is constrained to be the largest one, the remaining three
scales, being free parameters, can follow any hierarchy. In principle
six possible hierarchical patterns can be considered; however, since
lepton number is an intrinsic feature of seesaw models these
possibilities can be split in two generic scenarios:
\begin{enumerate}[I]
\item\label{scenarioI-fsl} The flavor symmetry related scales $M_\phi$
  and $v_F$ are larger than the number breaking scale.
\item\label{scenarioII-fsl} The flavor symmetry related scales $M_\phi$
  and $v_F$ are smaller than the number breaking scale.
\end{enumerate}
The scale at which leptogenesis takes place is intimately related with
the lepton number violating scale. Accordingly in scenarios
\ref{scenarioI-fsl} leptogenesis proceeds once the flavor symmetry is
already broken whereas in scenarios \ref{scenarioII-fsl} leptogenesis
takes place when the Lagrangian and the vacuum are still flavor
invariant i.e in the flavor symmetric phase. The former is considered in sections \ref{sec:single} and \ref{sec:both}, the latter cases
are the subject of this section.

From now on we will assume the Lagrangian and the vacuum to be
invariant under a flavor group $G_F$. The standard model leptons and
RH neutrinos, thus, belong to $G_F$ representations $R_a^{(X)}\sim
(X_1,\dots,X_m)$ (with $X=N,\ell,e$ and $a,b,c\dots$ denoting $G_F$
indices) in such a way that all the terms in (\ref{eq:seesaw-lag}) are
$G_F$ singlets. As can be seen in (\ref{eq:BmL-astmm-type-I}) a
vanishing $\epsilon_{N_1}$ implies in turn a vanishing $B-L$
asymmetry. Two conditions have to be satisfied in order to get
$\epsilon_{N_1}\neq 0$: ($i$) Mass splittings among the RH states,
otherwise the loop integrals arising from the vertex and wave function
corrections do not acquire an imaginary part; ($ii$) the matrix
$\pmb{m_D}^\dagger\pmb{m_D}$ must have non-zero and imaginary
off-diagonal elements. The first condition is satisfied if the RH
neutrinos belong to different $G_F$ representations (RH neutrinos
belonging to the same representation have a common universal mass). But
the second condition can never be achieved in the flavor symmetric phase: recovering the correct kinetic terms for the RH
neutrinos and lepton doublets requires
$R_a^{(N)*}R_b^{(N)}=\delta_{ab}$ and
$R_a^{(\ell)}R_b^{(\ell)*}=\delta_{ab}$, for the lepton doublets
transforming according to $\bar \ell\sim R^{(\ell)}_a$.  Taking the
scalar electroweak doublet as a $G_F$ singlet, the Yukawa coupling
matrix $\pmb{\lambda}$ is determined by the Clebsch-Gordan
coefficients arising from the contraction $R^{(\ell)}_aR^{(N)}_b$,
thus implying that the matrix
$\pmb{\lambda}^\dagger\pmb{\lambda}$ arises from the contractions
$R^{(N)*}_aR^{(\ell)}_bR^{(\ell)*}_cR^{(N)}_d=\delta_{ad}\delta_{bc}$
\cite{Sierra:2011vk}.

A non-vanishing $B-L$ asymmetry is possible only if new contributions
to the CP violating asymmetry exist ($\epsilon_{N_1}^\text{New}$)
i.e. if the flavons play a role, which they can do as propagating
states or virtually via loop corrections. In both cases the
kinematical constraint $M_\phi<M_N$ (where $M_N$ is the mass parameter
of the $R_a^{(N)}$ representation) must be guaranteed, as otherwise
either RH neutrino decays to flavons are kinematical forbidden or the
loop integral in which the flavons intervene can not acquire an
imaginary part. The flavor models one can envisage can be described by
a Lagrangian involving effective operators or models with ultraviolet
completions, regardless of the approach the presence of new energy
scales, different from that of lepton number violation, can have an
impact in the way leptogenesis takes place.  In \cite{Sierra:2011vk},
where the conditions for leptogenesis in the flavor symmetric phase
were established, an $A_4$ inspired model involving effective
operators was analyzed in full detail. In contrast,
\cite{AristizabalSierra:2007ur,AristizabalSierra:2009bh} discussed an
ultraviolet completed flavor toy model that we now discuss with the
purpose of illustrating the previous statements.

We will consider a setup inspired by $U(1)_X$ flavor models {\it \`a
  la} Froggatt-Nielsen \footnote{Leptogenesis in models based on the
  Froggatt-Nielsen mechanism have been studied in
  \cite{Cerdeno:2006ha}.}. Thus, in addition to the standard model
fields and RH neutrinos, the setup also contains vectorlike fermion
fields $F$ and a complex scalar field $S$ (flavon), all of them being
electroweak singlets.  With the horizontal charge assignment
$X(\ell,F)=+1$, $X(S)=-1$ and $X(H,N)=0$ the following Lagrangian can
be written
\begin{equation}
  \label{eq:U1-Lag}
  -{\cal  L}=\bar \ell\,\pmb{h}\,F\,H + \bar N\,\pmb{\lambda}\,F\,S
  +\frac{1}{2}\bar N^T\,C\,\pmb{M_N}\,N + \bar F\,\pmb{M_F}\,F\,.
\end{equation}
Here the Yukawa coupling matrices $\pmb{h}$ and $\pmb{\lambda}$ are
$3\times 3$ matrices in flavor space. The $U(1)_X$ symmetry is
spontaneously broken by the vacuum expectation value of the complex
scalar field, $\langle S\rangle=v_F$. In addition to the $U(1)_X$
symmetry the terms in the Lagrangian (\ref{eq:U1-Lag}) preserve a
global $U(1)$ symmetry with charge assignments $L(\ell,F,N)=+1$ and
$L(H,S)=0$. This symmetry is only broken by the RH Majorana mass term
and thus can be identified with lepton number. 
In this setup the scale $\Lambda$ corresponds to $M_F$, and $M_\phi$ to $M_S$. Since leptogenesis in the flavor symmetric phase requires
$M_N>v_F,M_S$ the following hierarchies follow $M_F>M_N>v_F,M_S$.
With $G_F$ being Abelian the standard contribution to the CP asymmetry
does not vanish, but due to the absence of the tree-level
coupling $\bar \ell N \tilde H$---enforced by the flavor charge
assignments---$\epsilon_{N_1}$ arises at the second loop-order, rendering
its value far below the one needed for successful leptogenesis
($\epsilon_{N_1}\gtrsim 10^{-6}$). Therefore, leptogenesis is viable only
if new contributions to the CP violating asymmetry are present.

\begin{figure}
  \centering
  \includegraphics[width=9cm,height=2.5cm]{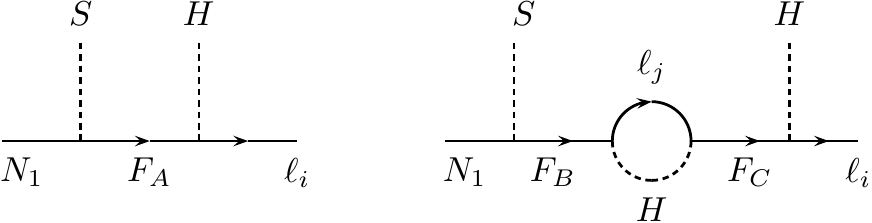}
  \caption{Tree-level and one-loop self-energy diagrams determining
    $\epsilon_{N_1}^{\text{(New)}\,\ell_i}$.}
  \label{fig:purely-flavored-lepto}
\end{figure}
With the couplings in (\ref{eq:U1-Lag}), and due to the kinematical
constraint $M_F>M_N, M_S$, RH neutrinos have three body decay modes,
$N_\alpha\to SH\ell_i$. So in this case the flavon $S$ intervenes in
the generation of the $B-L$ asymmetry as a propagating state. The
interference between the tree-level decay and the one-loop self-energy
correction diagrams shown in fig. \ref{fig:purely-flavored-lepto}
determine the new contribution to the flavored CP violating asymmetry,
which at leading order in the mass ratio $r_A=M_{N_1}^2/M_{F_A}^2$
reads \cite{AristizabalSierra:2007ur}
\begin{equation}
  \label{eq:new-contribution-CP}
  \epsilon^{(\text{New})\,\ell_i}_{N_1}=\frac{3}{128\pi}
  \frac{1}{(\pmb{\tilde\lambda}\pmb{\tilde\lambda})_{11}}
  \sum_j\mathbb{I}\text{m}
  \left[
    \left(
      \pmb{h}\,\pmb{\hat r}^2\pmb{h}^\dagger
    \right)_{ji}
    \tilde\lambda_{1j}\tilde\lambda_{1i}^*
  \right]\,,
\end{equation}
where $\pmb{\hat
  r}=\mbox{diag}(M_{N_1}^2/M_{F_1}^2,M_{N_1}^2/M_{F_2}^2,M_{N_1}^2/M_{F_3}^2)$
and the effective couplings $\pmb{\tilde \lambda}$ are defined as
\begin{equation}
  \label{eq:effective-couplings}
  \pmb{\tilde \lambda}=v_F\pmb{\lambda}\,\pmb{\hat M_F}^{-1}\,\pmb{h}^\dagger\,.
\end{equation}
The total CP violating asymmetry obtained from
(\ref{eq:new-contribution-CP}) by summing over the flavor indices 
vanishes
\begin{equation}
  \label{eq:new-contribution-CP-total}
  \epsilon_{N_1}^\text{New}=\sum_{i=e,\mu,\tau}
  \epsilon^{(\text{New})\,\ell_i}_{N_1}
  =\frac{3}{128\pi}
  \frac{1}{(\pmb{\tilde\lambda}\pmb{\tilde\lambda})_{11}}
  \sum_j\mathbb{I}\text{m}
  \left[
    \pmb{\tilde\lambda}\,
      \pmb{h}\,\pmb{\hat r}^2\pmb{h}^\dagger\,
      \pmb{\tilde\lambda}^\dagger
  \right]_{11}=0\,.
\end{equation}
Accordingly in the resulting scheme leptogenesis becomes possible only
via flavor dynamics and in that sense it is a purely flavored
leptogenesis realization
\cite{AristizabalSierra:2009bh,GonzalezGarcia:2009qd}.  Note that
since in this simple case $N_{2,3}$ are not involved in the loop
corrections the RH neutrino mass splittings are not relevant. Even if
they were relevant a mass splitting could always be accommodated due
to the Abelian nature of $G_F$. When $G_F$ is non-Abelian and the RH
neutrinos are placed in multiplets, as already stressed the mass
splittings can only be achieved if they belong to different
multiplets.
\section{Leptogenesis with flavor symmetries: type-I seesaw}
\label{sec:single}

The connection between flavor symmetry enforced Tribimaximal (TB)
mixing and leptogenesis was investigated by \cite{Jenkins:2008rb},
considering models based on $A_4$ and $Z_7 \rtimes Z_3$.  The
conclusion derived was that due to the specific construction those
models implement, the relevant quantity $\pmb{\cal M}\equiv \pmb{m_D}^{\dagger}
\;\pmb{m_D}$ is proportional to the identity matrix $\pmb{\mathbb{I}}$
and therefore the CP asymmetry must vanish at leading order (LO)
${\cal O}(\eta^0)$, with $\eta\equiv V/\Lambda$ and $V$ a generic
flavon vacuum expectation value $\langle\phi\rangle=V$. Importantly,
it was noted that there was a difference between having TB at low
energy accidentally (which allows leptogenesis to be viable) and TB
being enforced by a symmetry. Deviations from the exact mixing limit
were also considered and the magnitude of the CP asymmetry was
estimated as being connected to the magnitude of the next-to-leading
order (NLO), ${\cal O}(\eta^1)$, deviations of the mixing
angles. These conclusions were illustrated by considering the SUSY
model $A_4 \times Z_3$ of \cite{Altarelli:2005yx}.  In
\cite{Hagedorn:2009jy}, two specific $A_4$ models were carefully
studied (taking into account washout effects) in order to derive the
correlations between the deviation from the exact mixing limit and the
generation of leptonic asymmetries.  The existing collection of
particular cases was generalized into two model-independent results in
\cite{Bertuzzo:2009im, AristizabalSierra:2009ex}. Although the
conclusions of both generalizations are to some extent equivalent,
they rely on different assumptions and it is worth considering both in
detail. While \cite{Bertuzzo:2009im} is based on group theoretical
arguments, \cite{AristizabalSierra:2009ex} is based on general
arguments hinging explicitly on the absence of unnatural fine-tuning.

The group theoretical proof \cite{Bertuzzo:2009im} starts by
assuming invariance of the Lagrangian in (\ref{eq:seesaw-lag}) under a
generic flavor group $G_F$ in the limit $v_F=0$. Under this assumption
the Dirac and RH neutrino mass matrices must remain invariant under
$G_F$ transformations of $\ell$ and $N$, namely
\begin{equation}
  \label{eq:field-transformations}
  X\to \Omega_X(g)\,X\quad (\mbox{with}\;X=\ell,N)\,,
\end{equation}
where $\Omega_X(g)$ corresponds to unitary representations of the group
$G_F$ for the generic group element $g$.
Different conclusions can be derived depending on
whether the representations are irreducible or not:
\begin{itemize}
\item If 3 RH neutrinos are in a 3-dimensional irreducible
  representation the CP asymmetry vanishes at LO. Invariance of the
  Lagrangian implies the following equality
  \begin{equation}
    \label{eq:case1-irr-rep}
    \pmb{\cal M}= \Omega_N(g)^\dagger \pmb{\cal M} \Omega_N(g)\,.
  \end{equation}

  Since the irreducible representation is 3-dimensional $\Omega_N(g)$ is
  in general a non-diagonal matrix. Therefore as a direct consequence
  of this $\pmb{\cal M}$ is proportional to
  $\mathbb{I}$ so that the equality can be verified for any group
  element $g$. In general all the parameters in (\ref{eq:seesaw-lag})
  receive NLO corrections, from higher dimensional effective
  operators, and so do the total and flavored CP asymmetries.  At
  ${\cal O}(\eta^1)$ two cases can be identified:
  \begin{itemize}
  \item The loop-functions $f(\omega_\alpha)$ and $g(\omega_\alpha)$
    are independent of $\eta$: The flavored CP asymmetries
    $\epsilon_{N_\alpha}^{\ell_i}$ arise at ${\cal O}(\eta)$ as they
    have only one power of $\pmb{\cal M}$, and $(m_{D_{i
        \alpha}}^{\ast}\,m_{D_{i \beta}})$ needs not depend on $\eta$
    (the combination has flavor indices so the transformation
    properties of lepton doublets can be relevant).  With the sum over
    the lepton flavor index $i$ taken, the total CP asymmetry
    $\epsilon_{N_\alpha}$ depends on the square of
    $\pmb{\cal M}$ and it arises only at order
    $\eta^2$ (this is in agreement with the results of
    \cite{Jenkins:2008rb}).
  \item The non-Abelian symmetry produces degeneracies in the RH
    neutrino mass spectrum: there is an enhancement of one order
    in both asymmetries ($\epsilon_{N_\alpha}^{\ell_i} \sim {\cal
      O}(\eta^0)$, $\epsilon_{N_\alpha} \sim {\cal O}(\eta)$) due to the
    loop functions $f(\omega_\alpha)$ and $g(\omega_\alpha)$ having
    $\eta^{-1}$ dependence.
  \end{itemize}

\item If the RH neutrinos are in a reducible representation the
  conclusions do not follow so straightforwardly, but if the LO
  matrices $\pmb{M_N}$ and $\pmb{\cal M}$ are
  simultaneously diagonalizable then the same conclusions as in the
  case with irreducible representations apply. As a particular case,
  if the symmetry is Abelian its 1-dimensional representations are in
  general unable to make the asymmetry vanish, with the
  requirement that $\pmb{\cal M}$ is diagonal simultaneously with $\pmb{M_N}$ typically not being fulfilled.
\end{itemize}
The authors also investigated thoroughly a particular based in the
$A_4 \times Z_3 \times U(1)_{FN}$ model (\cite{Altarelli:2005yx}).


The general argument proof in \cite{AristizabalSierra:2009ex} starts
from an exact mixing scheme (in the form-diagonalizable sense
\cite{Low:2003dz}). The exact mixing is the outcome of a symmetry, not
accidental.  The proof relies fundamentally on the assumption that the
resulting effective light neutrino mass matrix can be diagonalized by
a special unitary matrix that does not depend on relationships between
the parameters that govern the masses. For definiteness the TB mixing
was considered:
\begin{equation}
  \label{eq:diagonalization}
  \pmb{\hat{m}_\nu}=\pmb{D}\,\pmb{U_{\text{TB}}}^T\, 
  \pmb{m_\nu^\text{eff}}\,\pmb{U_{\text{TB}}}\,\pmb{D}\,,
\end{equation}
$\pmb{D}$, defined in section \ref{sec:leptogenesis-type-I}, has the
low-energy Majorana phases and $\pmb{U_{\text{TB}}}$ is the PMNS
matrix with the corresponding TB values for the mixing angles. The Dirac and RH neutrino mass matrix are diagonalized
according to
\begin{equation}
\label{def}
\begin{array}{rcl}
\pmb{\hat{m}_D}&=&\pmb{U_L}^\dagger  \, \pmb{m_D}\, \pmb{U_R} \,,\\
\pmb{\hat{M}_N}&=&\pmb{V_R}^T \, \pmb{M_N}\, \pmb{V_R} \,,
\end{array}
\end{equation}
with $\pmb{U_{L,R}}$ and $\pmb{V_R}$ unitary matrices. Then, from the
seesaw formula we can write:
\begin{equation}
  \label{ss}
  \pmb{m_\nu^\text{eff}}= - \pmb{U_L}\,\pmb{\hat{m}_D} \, 
  \left(\pmb{U_R}^\dagger\,\pmb{V_R}\right)\, 
  \pmb{\hat{M}_N}^{-1}\, 
  \left(\pmb{V_R}^T \pmb{U_R}^*\right)\,
  \pmb{\hat{m}_D}\,\pmb{U_L}^T\,.
\end{equation}
We assume $\pmb{m_\nu^{\text{eff}}}$ is diagonalized by
the mixing scheme without special relationships between masses:
\begin{equation}
  \pmb{\hat{m}_\nu}= -  \pmb{D}\,
  \left(\pmb{U_{\text{TB}}}^T \pmb{U_L}\right)\,
  \pmb{\hat{m}_D} \, \left(\pmb{U_R}^\dagger\,\pmb{V_R}\right)\,
  \pmb{\hat{M}_N}^{-1}\, 
  \left(\pmb{V_R}^T \pmb{U_R}^*\right)\,
  \pmb{\hat{m}_D}\,
  \left(\pmb{U_L}^T \pmb{U_{\text{TB}}}\right)\,\pmb{D}\,,
\end{equation}
therefore the matrix on the left-hand side is diagonal (denoted by the
hat), which then implies that the combinations of matrices appearing
on the right-hand side of the equation, $\left(\pmb{U_{\text{TB}}}^T
  \pmb{U_L}\right)$, $\left(\pmb{U_R}^\dagger\,\pmb{V_R}\right)$ and
conversely $\left(\pmb{V_R}^T \pmb{U_R}^*\right)$, $\left(\pmb{U_L}^T
  \pmb{U_{\text{TB}}}\right)$ should also be diagonal (up to
orthogonal rotations in case of degenerate eigenvalues, but this does
not alter the conclusion). Consider for simplicity a case without
degeneracies, and evaluate off-diagonal elements of the expression on
the right hand side: if the matrix combinations identified above were
not diagonal, then the off-diagonal elements of the right-hand side
will depend on combinations of the masses of RH neutrinos and Yukawa
couplings $\pmb{\pmb{\lambda}}$, which could only vanish for very
specific relations between them---which explicitly violates
form-diagonalizability. Therefore $\left(\pmb{U_{\text{TB}}}^T
  \pmb{U_L}\right)$, $\left(\pmb{U_R}^\dagger\,\pmb{V_R}\right)$
should indeed be diagonal up to orthogonal rotations of degenerate
eigenvalues.  Assuming no degeneracies this implies
\begin{equation}
  \label{eq:UL-and-UR}
  \pmb{U_L}=\pmb{U_{\text{TB}}}\,\pmb{\hat P_L}\,,
  \qquad
  \pmb{U_R}^\dagger=\pmb{\hat P_R}\,\pmb{V_R}^\dagger\,,
\end{equation}
with $\pmb{\hat
  P_{L,R}}=\mbox{diag}(e^{\text{i}\alpha_1^{L,R}},e^{\text{i}\alpha_2^{L,R}},
e^{\text{i}\alpha_3^{L,R}})$. These relations allow to fix the
structure of the Dirac mass matrix as
\begin{equation}
  \label{eq:Dirac-mass-matrix-secTBM}
  \pmb{m_D}=\pmb{U_{\text{TB}}}\,\pmb{\hat D}^*\pmb{\hat m_D}\,,
\end{equation}
that when compared with the Casas-Ibarra parametrization in
(\ref{eq:casas-ibarra}) leads to $\pmb{R}=\pmb{\hat
  m_\nu}^{-1/2}\,\pmb{\hat m_D}\,\pmb{\hat M_N}^{-1/2}$, showing that
$\pmb{R}$ is diagonal and real. As the total CP asymmetry can be
expressed as:
\begin{equation}
  \label{eq:cp-asymm-CI}
  \epsilon_{N_\alpha} = -\frac{3 M_{N_\alpha}}{8 \pi v^2}
  \frac{{\mathbb I}\mbox{m}
    \left[\sum_i m_{\nu_i}^2 R_{i \alpha}^2\right]}
  {\sum_i m_{\nu_i} |R_{i \alpha}|^2}\,,
\end{equation}
then the asymmetry must vanish.  Alternatively one can consider the
following: $(\pmb{U_R}^\dagger\,\pmb{V_R})$ is diagonal from our
assumption, this means that the basis where $\pmb{m_D}$ is diagonal
and the basis where $\pmb{M_N}$ is diagonal have a special
relationship (this is often denoted as form-dominance
\cite{Chen:2009um} and is essentially also the requirement outlined in
the group theoretical approach of \cite{Bertuzzo:2009im} in the case of reducible representations).
We can simply start with the diagonal basis of $\pmb{m_D}$, use
$\pmb{U_R}$ to bring it to the general basis, $\pmb{V_R}$ to bring it
to the basis of diagonal $\pmb{M_N}$ and see that in that basis
$\pmb{m_D}$ is essentially $\pmb{U_{\text{TB}}}\;\pmb{\hat{m}_D}$---its
columns are the eigenvectors of the mixing scheme. Naturally when
$\pmb{m_D}^{\dagger} \pmb{m_D}$ is taken the mixing cancels out and
the relevant quantity for the asymmetry is diagonal (consistently with
\cite{Jenkins:2008rb, Bertuzzo:2009im}).  Although the example uses TB
mixing for definiteness, it should be stressed that any exact mixing
scheme enforced by a symmetry leads to the same conclusion. The paper
also looked into several particular cases of TB mixing, dividing them
into classes of models according to the structure of $\pmb{m_D}$ and
$\pmb{M_N}$. The structure of NLO contributions was considered
explicitly with expansions around the the LO values, leading to:
\begin{align}
  \label{NLO-corrections}
  \pmb{m_D}^{\prime\dagger} \pmb{m_D}^{\prime}&=
  \pmb{m_D}^{\dagger} \pmb{m_D} +
  \pmb{m_D}^{\dag}
  \left(
    \pmb{U_\ell^{(1)}}^\dag\pmb{m_D} +
    \pmb{U_L}\pmb{U_L^{(1)}}\pmb{\hat{m}_D} \pmb{U_R}^{\dag}\pmb{V_R} +
    \pmb{U_L}\pmb{\hat{m}_D}^\prime \pmb{U_R}^\dag \pmb{V_R} +
  \right.\nonumber\\
   & + \left.
    \pmb{U_L}\pmb{\hat{m}_D}\pmb{U_R^{(1)}}^\dag\pmb{U_R}^\dag\pmb{V_R} +
    \pmb{m_D} \pmb{V_R^{(1)}}\right) + \mathrm{h.c.}\,.
\end{align}
The superscript (1) refers to those quantities corrected by
NLO contributions and $\pmb{U_\ell}$ diagonalizes the
charged lepton mass matrix (we started on the basis where it is diagonal at LO, but it becomes non-diagonal after NLO
corrections are introduced).
The $A_4$ model of \cite{Lin:2009ic} was used to illustrate the
conclusions and to highlight how it can be possible to link low and
high-energy CP violation parameters.  Finally it was noted that with
added degrees of freedom (such as from having type-II seesaw) it would
be possible to generate an asymmetry even while remaining in the exact
mixing limit.

Not long after these two important generalizations, further
results were presented by \cite{Felipe:2009rr} and \cite{Choubey:2010vs}, clarifying some points that
we summarize very briefly here.
Assuming that the symmetries of the mass matrices involved
in type-I are residual symmetries of the Lagrangian,
\cite{Felipe:2009rr} shows that $\pmb{{\cal M}}$ is diagonal and
therefore the asymmetry vanishes. They also consider the exact mixing schemes
so characteristic of models with flavor symmetries and connect that
requirement with their assumption: if the effective neutrino mass
matrix has nonzero determinant, then the Lagrangian contains the
maximal residual symmetry (that of the mass matrices) and so
leptogenesis can not proceed at LO and in fact even when the
determinant vanishes, $\epsilon_{N_\alpha}$ is still zero at LO.
The implication of form-dominance \cite{Chen:2009um} on the
Casas-Ibarra matrix $\pmb{R}$ is considered in detail in
\cite{Choubey:2010vs}: the vanishing CP
asymmetry is not particular to TB. Rather, exact
mixing schemes enforced by symmetries are a particular case of
form-dominance \cite{Chen:2009um}. The main conclusions are that form-dominance by itself is sufficient
to make the CP asymmetry vanish and that it is possible to violate form-dominance softly without perturbing the mixing.
The cases considered earlier in \cite{Antusch:2006cw} were summarised, and they exemplify very clearly the separation between TB and form-dominance.

Before concluding this section it is important to stress that
corrections to the exact mixing scheme are typically expected at NLO
as is explicitly considered in the literature (see
e.g. \cite{Altarelli:2010gt}). An exception which does preserve exact
mixing is one of the renormalizable UV complete models in
\cite{Varzielas:2010mp} (due to the lack of certain
messengers). Furthermore, in \cite{Cooper:2011rh} it is shown that RG
corrections can also provide the deviations necessary to lift the
vanishing CP asymmetries.

\section{Leptogenesis with flavor symmetries: type-I and II seesaws}
\label{sec:both}

Recently, a model-independent analysis in the style of
\cite{AristizabalSierra:2009ex} considered cases with both type-I and
type-II seesaw \cite{AristizabalSierra:2011ab}. Flavor models limit themselves to type-I and/or II seesaws with few exceptions (e.g. \cite{Bazzocchi:2009qg}). As noted in
\cite{AristizabalSierra:2009ex}, in general the CP asymmetries
involving the additional degrees of freedom can be non-vanishing even
in the exact mixing limit and \cite{AristizabalSierra:2011ab}
considered the framework with both seesaw types in detail. It was
shown that non-vanishing CP asymmetries depend on the existence of
repeated eigenvectors across the seesaw types. The main point is the
following: leptogenesis can become viable through the
CP asymmetries in which the triplets intervene
i.e. $\epsilon_{N_\alpha}^\Delta$ or $\epsilon_\Delta$ (see
eqs. (\ref{eq:cp-asymm-triplet}) and (\ref{eq:triplet-CPasymmetry})),
depending---of course---on whether it proceeds via RH neutrino or
scalar triplet dynamics (or both as in the case treated in
sec. \ref{sec:leptogenesis-type-II}). Both CP asymmetries depend on
the imaginary parts of
\begin{equation}
  \label{eq:matrix-involved-type-i-ii}
  \pmb{{\cal Y}}\,\mu=\pmb{m_D}\,\pmb{Y}^*\,\pmb{m_D}^T\,\mu\,.
\end{equation}
Since the parameter $\mu$ is in general complex, and the presence of
$G_F$ does not allow a definitive statement about its
phase, the CP asymmetries are non-vanishing even if the matrix
$\pmb{\cal Y}$ turns out to be real. Vanishing $\pmb{\cal Y}$,
however, implies $\epsilon_{N_\alpha}^\Delta,\epsilon_\Delta=0$. In
that sense the quantity to be analyzed is $\pmb{\cal Y}$.

Definitive conclusions about this matrix can be made by writing the
effective light neutrino mass matrix as an outer product of the
eigenvectors of the assumed mixing scheme \footnote{These eigenvectors
  are determined by the column vectors of the PMNS matrix for a fixed
  mixing pattern.} and its mass eigenvalues:
\begin{equation}
  \label{eq:mass-matrix-eff-i-plus-ii}
  \pmb{m_\nu^\text{eff}}=\sum_{i=e,\mu,\tau}=m_{\nu_i}\,
  \pmb{v_i}\otimes\pmb{v_i}\,.
\end{equation}
According to (\ref{eq:eff-mass-matrix}) the eigenvectors
come from the contributions of type-I and/or type-II seesaws:
\begin{equation}
  \label{eq:sesaw-mm-outer}
    \pmb{m_\nu^X} = \sum_{i=1,2,3}m_{\nu_i}^X 
  \pmb{v_i}\otimes\pmb{v_i}\qquad (X=I,II)\,.
\end{equation}
This decomposition is based on the assumption that both
$\pmb{m_\nu^I}$ and $\pmb{m_\nu^{II}}$ are diagonalized by the PMNS
matrix fixed by the assumed mixing scheme. This needs not be the case
but if it is not then somehow a contribution that is incompatible with
the mixing scheme is present in both seesaw types in just the correct
quantities to cancel each other out (which amounts to unrealistic
fine-tuning given the separate physical degrees of freedom involved).
With the decomposition in (\ref{eq:sesaw-mm-outer}) we classify
the possible models:
\begin{enumerate}[A)]
\item\label{g-models}
  \underline{General models}: The eigenvectors
  $\pmb{v_i}$, defining the effective light neutrino mass matrix, stem
  from both type-I and type-II contributions. Note that in this case
  in addition to the pieces involving the eigenvectors $\pmb{v_i}$
  each (or only one) seesaw contribution may involve also the identity
  matrix $\mathbb{I}$.
\item\label{int-models}
  \underline{Intermediate models}: The
  eigenvectors $\pmb{v_i}$ entirely arise from either type-I or type-II contributions.
\item\label{min-models}
  \underline{Minimal models}: Two eigenvectors
  $\pmb{v_i}$ stem from the type-I (type-II) contributions and the
  third one $\pmb{v_k}$ (with $\pmb{v_i}\cdot\pmb{v_k}=0$) from type-II (type-I).
\end{enumerate}
Note that being able to parametrize each seesaw contribution with
these eigenvectors does not mean they are all explicitly present. A common scenario can be the explicit presence of only a
single eigenvector in a given seesaw type in either cases
\ref{int-models} or \ref{min-models} (with at least one more eigenvector present in the other seesaw).
Another relevant observation is that the underlying symmetry may
be arranging structures which can be reparametrized in terms of
the eigenvectors, meaning one does not necessarily need separate
physical degrees of freedom to have more than one eigenvector
represented---see e.g. \cite{deMedeirosVarzielas:2011tp} where the
$\mu-\tau$ structure
\begin{equation}
  P=\begin{pmatrix}
    1 & 0 & 0 \\
    0 & 0 & 1 \\
    0 & 1 & 0
  \end{pmatrix}
\end{equation}
arises directly from specific discrete groups---as a TB-compatible
contribution it can be expressed in terms of the TB
eigenvectors 
as explicitly seen with $b=2a$ and $c=-3a$ in the parametrization:
\begin{equation}
  \label{eq:tbm-light-mm}
    \pmb{m_\nu^{\text{eff}}}=\pmb{m_\nu^I} + \pmb{m_\nu^{II}}
    = 
    \begin{pmatrix}
      4a + b & -2a +b & -2a + b\\
      \cdot & a+b+c & a+b-c\\
      \cdot & \cdot & a+b+c
    \end{pmatrix}\,.
\end{equation}

With the models classified according to the eigenvectors of
their mixing scheme, we can determine the structures of $\pmb{m_D}$, 
$\pmb{Y}$ and then also $\pmb{{\cal
    Y}}$ (see \cite{AristizabalSierra:2011ab} for details). However, even without determining explicitly these
structures, it can be realized from the definitions in
\ref{g-models}, \ref{int-models} and \ref{min-models} that vanishing
$\pmb{{\cal Y}}$ occurs only when $\pmb{m_D}$ and $\pmb{Y}$ are
orthogonal, and in principle this happens only in models of type
\ref{min-models}: in the other cases, the presence of
the scalar triplet degrees of freedom allows the generation of the baryon asymmetry via leptogenesis even in
the limit of an exact mixing pattern (in agreement
with what was suggested in \cite{AristizabalSierra:2009ex}).

\begin{figure}
  \centering
  \includegraphics[width=10.5cm,height=7cm]{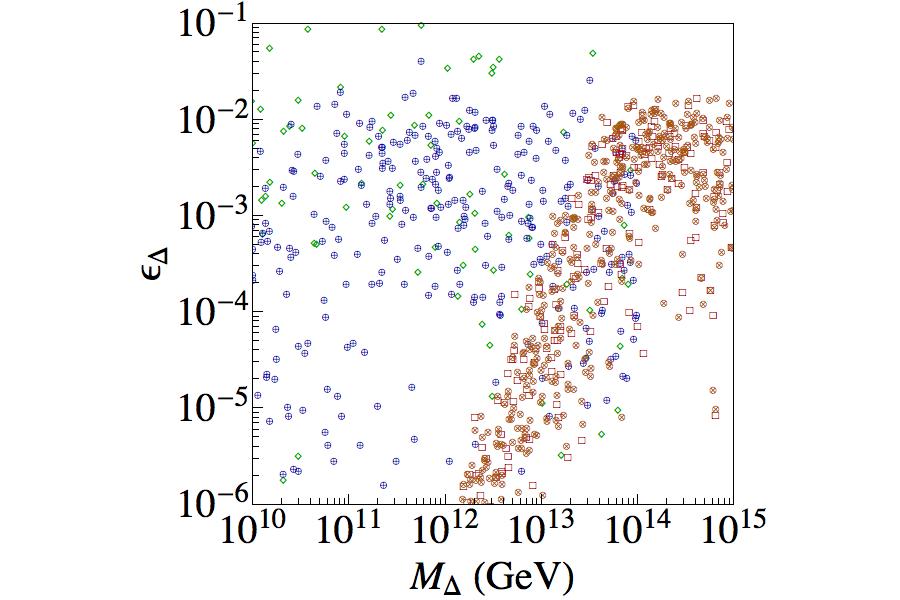}
  \caption{\it $\epsilon_\Delta$ as function of $M_\Delta$. Red
    squares and orange crosses for normal and inverted hierarchy of a
    specific 3-parameter predictive case. Green squares and blue
    crosses for normal and inverted hierarchy of the general
    8-parameter case \cite{AristizabalSierra:2011ab}.}
  \label{fig1}
\end{figure}

Having identified models where leptogenesis becomes viable in the exact mixing limit, the obvious step is to study those in
which the constraints enforced by $G_F$ allow the CP asymmetry
to be constrained by the low-energy data. There are in general 6 observables: 3 light neutrino masses and
3 CP phases (only 2 Majorana phases in TB mixing). Thus,
models involving more than 6 parameters barely allow to make any
statement about the asymmetry. The most general models in
\ref{g-models} are within that class, with 8 complex
parameters. Models in which the asymmetries
$\epsilon_{N_\alpha}^\Delta,\epsilon_\Delta$ can be constrained by
the low-energy data fall within classes
\ref{int-models} or \ref{min-models}. One
can add a contribution proportional to $\mathbb{I}$ to either
(or both) seesaw types, and any such contribution counts as all
(and any) 3 eigenvectors, so the quantity $\mathfrak{I}\mbox{m}[\pmb{\cal
  Y}]$ can be expressed in terms of the combinations of parameters
defining the quantities $\pmb{m_{\nu_i}^{I,II}}$. Denoting them
as $a_0^X$ and $a_i^X$ ($X=I,II$) for $\mathbb{I}$
and the eigenvectors contributions respectively, it turns out that
\begin{equation}
  \label{eq:explicit}
  \mathfrak{I}\mbox{m}\left[\pmb{\cal Y}\right]=
  \mathfrak{I}\mbox{m}
  \left[ (a_0^I a_0^{II \star}) 
    + \sum_i \left( a_i^I a_i^{II \star} \right) 
    + \left(\sum_i a_i^I \right) a_0^{II \star} 
    + a_0^I \left(\sum_i a_i^{II \star}\right) \right]\,.
\end{equation}
In particular for the class of models discussed in item \ref{min-models}
with only two eigenvectors $\pmb{v_i}$ stemming from type-I (type-II) and 
$\mathbb{I}$ from type-II (type-I) we have
\begin{equation}
  \label{eq:explicit-particular}
  \mathfrak{I}\mbox{m}\left[\pmb{\cal Y}\right]=
  \mathfrak{I}\mbox{m}
  \Bigg[(a_0^X)^\star \sum_{\substack{i<j\\j=2,3}} a_i^Y\Bigg]\,,
\end{equation}
with $X=I$ and $Y=II$ or vice versa. The parameters of these models are only 3 and can be well restricted by means of the solar and
atmospheric squared mass differences \cite{Schwetz:2011zk} yielding
tight constraints on the CP asymmetries. Figure \ref{fig1} shows $\epsilon_\Delta$ in models for which two eigenvectors
originate from type-I and the contribution from type-II is
proportional to $\mathbb{I}$, assuming a TB mixing pattern.  For
comparison we have also included the results for the general cases
discussed in \ref{g-models} involving contributions proportional to
the $\mathbb{I}$ in type-I and II. The scatter plot
was obtained by randomly scanning the parameters defining the neutrino
masses and selecting those points that lead to solar and
atmospheric squared mass differences within the experimental range. Figure \ref{fig1} shows that in general models,
even in the limit of an exact mixing pattern no statement
about the CP asymmetry can be established. In contrast, in the
simplified model considered, specific values of the CP asymmetry
require somehow specific ranges for the triplet mass.

\section{Conclusions}
\label{sec:conclusions}
From a general perspective the problem of studying leptogenesis in the
presence of flavor symmetries $G_F$ depends on whether the lepton
number breaking scale $\Lambda_L$ is above or below the scales
involved in $G_F$ (flavor breaking and flavon scales, denoted
generically $v_F$ and $M_\phi$).  In the flavor symmetric
phase, defined as a scenario in which $\Lambda_L>v_F,M_\phi$, as
described in section \ref{sec:sym}, flavons must play a relevant role
in the generation of the lepton asymmetry either as propagating or
loop intermediate states. Indeed it turns out that the presence of
these states apart from rendering leptogenesis viable can change the
conventional picture by e.g. leading to models in which leptogenesis
proceeds entirely via lepton flavor effects\footnote{The viability of
these models depends on whether lepton flavor equilibrating effects
can be circumvented \cite{AristizabalSierra:2009mq}.}.

In the case of type-I seesaw in the flavor broken phase,defined as a
scenario where $v_F,M_\phi>\Lambda_L$, the model-independent
conclusion reviewed in section \ref{sec:single} is that CP asymmetries
vanish in the exact mixing limit enforced by flavor symmetries. This is not an intrinsic feature of the exact
mixing, and this result can be attributed to the property of form
dominance in the neutrino mass matrices.  Within the scenario of
type-I seesaw with symmetry enforced mixing, interesting correlations
between low energy observables (mixing angles and CP phases) and
high-energy parameters (CP asymmetries) can be present when there is
departure from the exact mixing limit. It is possible even in a
model-independent context to identify rather generally the order of
magnitudes associated with a small parameter responsible for the
mixing deviations.

When other degrees of freedom that can contribute to leptogenesis are
added, such as those associated with type-II seesaw, the above
conclusions need not apply. Section \ref{sec:both} considers specifically the interplay between
type-I and II, where it is possible to conclude that the associated asymmetries still vanish in
special cases. Classifying these hybrid scenarios according to
the eigenvectors of the exact mixing is helpful, and even
without departure from exact mixing leptogenesis can occur whenever
eigenvectors are repeated across the two seesaw
types---with contributions proportional to $\mathbb{I}$ counting as
any and all eigenvectors.

Finally, we note that in accordance with section \ref{sec:sym}, in the
flavor symmetric phase it is possible to have non-vanishing asymmetry
originating just from type-I seesaw while the type-II asymmetries
vanish due to orthogonality of the eigenvectors. For this to occur
there must be a specific hierarchy of scales so that the RH neutrinos
decay in the flavor symmetric phase, avoiding the results described in
section \ref{sec:single}, while $\Delta$ decays in the broken phase
with vanishing contributions as described in section \ref{sec:both}.

\section*{Acknowledgement}
We specially thank Federica Bazzocchi for helpful discussions. DAS
also wants to acknowledge Marta Losada, Luis Alfredo Mu\~noz, Jernej
Kamenik and Miha Nemev\v{s}ek for the enjoyable collaboration on the
subjects discussed here. Special thanks to Enrico Nardi for the always
enlightening leptogenesis discussions. DAS is supported by a Belgian
FNRS postdoctoral fellowship.  IdMV is supported by DFG grant PA
803/6-1 and partially through PTDC/FIS/098188/2008.
\appendix
\section{Conventions and notation}
\label{sec:conventions}
In this appendix we collect the equations used in the calculations
discussed in section \ref{sec:lepto-gen}. We start by specifying
well known statistical and cosmological quantities.
\subsection{Equilibrium distributions and Cosmological quantities}
\label{sec:eq-dis-cosmo}
All the results presented in this short review were done using
Maxwell-Boltzmann distribution functions. For type-I and type-III
seesaws the equilibrium number densities read
\begin{equation}
  \label{eq:equilibrim-num-densities}
  n^{\text{Eq}}_{\ell,H}(z)=\frac{2M^3}{\pi^2\;z^3}\,,
  \quad
  n^{\text{Eq}}_{N_1}(z) =\frac{M^3_{N_1}}{\pi^2}\frac{K_2(z)}{z}\,,
  \quad
  n^{\text{Eq}}_{X}(z) =\frac{3M^3_{X}}{2\pi^2}\frac{K_2(z)}{z}
  \quad(\mbox{with}\;\; X=\Delta,T_\alpha)\,.
\end{equation}  
Here $K_2(z)$ is the second-order modified Bessel function of the
second-type and $z\equiv M/T$ where $M$ can refer to
$M_{N_{1},\Delta,T_\alpha}$ depending on the considered case (this also
applies for $n^{\text{Eq}}_{\ell,H}(z)$).  For the type-II scenario
discussed in \ref{sec:leptogenesis-type-II} the $N_1$ equilibrium
number density is given by
\begin{equation}
  \label{eq:N1-equilibrium-number-dens-type-II}
  n^{\text{Eq}}_{N_1}(z) =\frac{M^3_\Delta}{\pi^2}r^2\frac{K_2(rz)}{z}\,,
\end{equation}
with $r=M_{N_1}/M_\Delta$. The energy density $\rho(z)$ and pressure
$p(z)$ become
\begin{equation}
  \label{eq:energy-dens}
  \rho(z)=\frac{3 M^4}{z^4\pi^2} g_*\,,\quad
  p(z)=\frac{M^4}{z^4\pi^2} g_*
\end{equation}
where $g_*=\sum_{i=\text{All species}} g_i$ is the number of standard
model relativistic degrees of freedom (118 for $T\gg$ 300
GeV). Accordingly, the expansion rate of the Universe and entropy
density can be written as
\begin{equation}
  \label{eq:HandS}
  H(z)=\sqrt{\frac{8g_*}{\pi}}\frac{M^2}{M_{\text{Planck}}}
  \frac{1}{z^2}\,,
  \quad
  s(z)=\frac{4 M^3}{z^3\pi^2} g_*\,.
\end{equation}
\subsection{Formal solutions of the kinetic equations}
\label{sec:formal-sol}
In the type-I and III seesaw cases the integration of the differential
equations accounting for the evolution of the $B-L$ asymmetry leads to
\begin{equation}
  \label{eq:BmL-appendix-typeI-III}
  Y_{\Delta_{B-L}}(z)=-n\times\epsilon_X\,Y_X^\text{Eq}(z_0)\eta(z)\,,
\end{equation}
where $X=N_1,T_1$ and $n=1,3$ depending on whether the decaying state
is the singlet or the triplet. Assuming a vanishing initial asymmetry
($Y_{\Delta_{B-L}}^\text{(In)}=0$) the efficiency function can be
written as
\begin{equation}
  \label{eq:eff-type-I-III}
    \eta(z)=\frac{1}{Y_{T_\alpha}^{\text{Eq}}(z_0)}
  \int_{z_0}^{z}\;Q_X(z')\frac{dY_{X}(z')}{dz'}
  e^{-\int_{z'}^z dz'' P_{X}(z'')}\;,
\end{equation}
with the functions $Q_X(z),P_X(z)$ given by
\begin{equation}
  \label{eq:integrand-efficiencyappendix}
  Q_{N_1}(z)=1\,,\quad
  Q_{T_1}(z)=\frac{\gamma_{D_{T_1}}}{\gamma_{D_{T_1}}+2\gamma_{A}}\,,\quad
  P_{N_1,T_1}(z)=\frac{1}{2 Y^{\text{Eq}}_\ell(z)}
  \frac{\gamma_{D_{N_1,T_1}}(z)}{s(z) H(z) z}\,.
\end{equation}
Freeze-out of the asymmetry is at $z=z_f$ with $z_0\ll
z_f$. The efficiency factor is determined by $\eta=\eta(z_f)$.

The case for type-II is more involved but the kinetic equation for the
$L$ asymmetry in (\ref{eq:BEQ-type-II}) can still be formally
integrated \cite{AristizabalSierra:2011ab}. Again, assuming an initial
vanishing $L$ asymmetry, we get
\begin{equation}
  \label{eq:BmL-formal-sol}
  Y_{\Delta_L}(z)=\int_{z_i}^z\,dz'\,Q(z')\,e^{-\int_{z'}^z\,dz''\,P(z'')}\,,
\end{equation}
with the functions $Q(z)=Q^I(z) + Q^{II}(z)$ and $P(z)$ given by
\begin{align}
  \label{eq:PandQ-functions1}
  Q^I(z)&=\frac{1}{s(z)H(z)z}
  \left\{
    \left[
      (y_{N_1}(z)-1)\epsilon_{N_1}^\text{tot}-y_{\Delta_\Delta}^H(z)
    \right]\gamma_{D_{N_1}}(z)
  \right\}\,,\\
  \label{eq:PandQ-functions2}
  Q^{II}(z)&=\frac{3}{s(z)H(z)z}
  \left\{
    \left[
      (y_\Sigma(z) - 1)\epsilon_\Delta - 2 K_\ell\,y_{\Delta_\Delta}(z)
    \right]\gamma_{D_\Delta}(z)
  \right\}\,,\\
  \label{eq:PandQ-functions3}
  P(z)&=\frac{1}{s(z)H(z)z}
  \left[
    \frac{1}{Y^\text{Eq}_\ell}
    \left(
      \gamma_{D_{N_1}}(z) + 2\,K_\ell\,\gamma_{D_\Delta}(z)
    \right)
  \right]\,.
\end{align}
Note that in $Q^{II}(z)$ we have included a factor of 3 coming from
the $SU(2)$ physical degrees of freedom of the triplet. By factorizing either
$\epsilon_{N_1}^\text{tot}$ or $\epsilon_\Delta$ from the functions
$Q^{I,II}(z)$ and normalizing to $Y^\text{Eq}_\text{tot}\equiv
Y^\text{Eq}_\text{tot}(z\to0)=
Y^\text{Eq}_{N_1}(z)+Y^\text{Eq}_\Sigma(z)|_{z\to 0}$ the $L$
asymmetry in (\ref{eq:BmL-formal-sol}) can be written in terms of
efficiency functions that depend on the dynamics of the scalar triplet
and the fermionic singlet as done in eq. (\ref{eq:efficiencies}).
\subsection{Reduced cross sections for triplet scalar and fermion}
\label{sec:reduced-cross-secion}
The reduced cross section for the scalar electroweak triplet involves
the $s$-channel processes $\pmb{\Delta}\pmb{\Delta}\to F\bar F, AA,
HH$ ($F$ and $A$ stand for standard model fermions and $SU(2)\times
U(1)$ gauge bosons respectively), $t$ and $u$ channel triplet mediated
processes $\pmb{\Delta}\pmb{\Delta}\to AA$ and the ``quartic'' process
$\pmb{\Delta}\pmb{\Delta}\to AA$. In powers of the kinematic factor
$\omega(x)=\sqrt{1-4/x}$ (with $x=M_\Delta^2/s$) it can be split in
three pieces \cite{Hambye:2005tk}:
\begin{align}
  \label{eq:reduced-cross-section}
  \widehat \sigma_1(x)&=\frac{1}{\pi}
  \left[
    g^4\left(5 + \frac{34}{x}\right)
    +
    \frac{3}{2}g'^4\left(1 + \frac{4}{x}\right)
  \right]\omega(x)\,,
  \nonumber\\
  \widehat \sigma_2(x)&=\frac{1}{8\pi}
  \left(
    25 g^4 + \frac{41}{2}g'^4
  \right)\omega(x)^3\,,
  \nonumber\\
  \widehat \sigma_3(x)&=\frac{6}{\pi x^2}
  \left[
    4 g^4 (x-1) + g'^4 (x-2)
  \right]\ln\left[
    \frac{1+\omega(x)}{1-\omega(x)}
  \right]
  \,,
\end{align}
with $\widehat\sigma_A(x)=\sum_{i=1}^3\widehat \sigma_i(x)$ and $g,
g'$ the $SU(2)$ and $U(1)$ gauge couplings.

For the fermion $SU(2)$ triplet the reduced cross sections involves
the gauge boson mediated $s$-channel processes $T_\alpha
T_\alpha\leftrightarrow \ell\bar\ell$ and $T_\alpha
T_\alpha\leftrightarrow q\bar q$ and the $t$ and $u$-channel triplet
mediated process $T_\alpha T_\alpha\leftrightarrow A_\mu A^\mu$. The full
result where now $x=M_{T_1}^2/s$, reads \cite{Hambye:2003rt}:
\begin{equation}
  \label{eq:red-cross-sec}
  \widehat \sigma_A(x)=
  \frac{6g^4}{\pi}\left(1+\frac{2}{x}\right)\omega(x)
  +
  \frac{2g^4}{\pi}\left[
    3\left(
      1 + \frac{4}{x} - \frac{4}{x^2}
    \right)
    \log
    \left(
      \frac{1+\omega(x)}{1-\omega(x)}
    \right)
    -\left(
      4 + \frac{17}{x}
    \right)\omega(x)
  \right]\,.
\end{equation}

\bibliographystyle{JHEP}
\bibliography{refs}

\providecommand{\href}[2]{#2}\begingroup\raggedright\begin{thebibliography}{10}

\bibitem{Sakharov:1967dj}
A.~Sakharov, {\it {Violation of CP Invariance, C Asymmetry, and Baryon
  Asymmetry of the Universe}},  {\em Pisma Zh.Eksp.Teor.Fiz.} {\bf 5} (1967)
  32--35. Reprinted in *Kolb, E.W. (ed.), Turner, M.S. (ed.): The early
  universe* 371-373, and in *Lindley, D. (ed.) et al.: Cosmology and particle
  physics* 106-109, and in Sov. Phys. Usp. 34 (1991) 392-393 [Usp. Fiz. Nauk
  161 (1991) No. 5 61-64].

\bibitem{Minkowski:1977sc}
P.~Minkowski, {\it {$\mu\to e\gamma$ at a Rate of One Out of 1-Billion $\mu$
  Decays?}},  {\em Phys.Lett.} {\bf B67} (1977) 421.

\bibitem{Yanagida:1979as}
T.~Yanagida, {\it {Horizontal gauge symmetry and masses of neutrinos}},  in
  {\em Proceedings of the Workshop on Unified Theories and Baryon Number in the
  Universe : National Laboratory for High Energy Physics (KEK)}, Tsukuba: KEK,
  1979.

\bibitem{Glashow:1979nm}
S.~Glashow, {\it {The future of elementary particle physics}},  {\em NATO Adv.
  Study Inst.Ser.B Phys.} {\bf 59} (1980) 687. Preliminary version given at
  Colloquium in Honor of A. Visconti, Marseille-Luminy Univ., Jul 1979.

\bibitem{GellMann:1980vs}
M.~Gell-Mann, P.~Ramond, and R.~Slansky, {\it {Complex spinors and unified
  theories}},  in {\em Supergravity Workshop}, Elsevier Science Ltd, 1980.
\newblock Print-80-0576 (CERN).

\bibitem{Mohapatra:1979ia}
R.~N. Mohapatra and G.~Senjanovic, {\it {Neutrino mass and spontaneous parity
  nonconservation}},  {\em Phys. Rev. Lett.} {\bf 44} (1980) 912.

\bibitem{Schechter:1980gr}
J.~Schechter and J.~W.~F. Valle, {\it {Neutrino Masses in $SU(2) \times U(1)$
  Theories}},  {\em Phys. Rev.} {\bf D22} (1980) 2227.

\bibitem{Lazarides:1980nt}
G.~Lazarides, Q.~Shafi, and C.~Wetterich, {\it {Proton Lifetime and Fermion
  Masses in an $SO(10)$ Model}},  {\em Nucl.Phys.} {\bf B181} (1981) 287.

\bibitem{Mohapatra:1980yp}
R.~N. Mohapatra and G.~Senjanovic, {\it {Neutrino Masses and Mixings in Gauge
  Models with Spontaneous Parity Violation}},  {\em Phys.Rev.} {\bf D23} (1981)
  165.

\bibitem{Wetterich:1981bx}
C.~Wetterich, {\it {Neutrino Masses and the Scale of $B-L$ Violation}},  {\em
  Nucl.Phys.} {\bf B187} (1981) 343.

\bibitem{Foot:1988aq}
R.~Foot, H.~Lew, X.~He, and G.~C. Joshi, {\it {Seesaw neutrino masses induced
  by a triplet of leptons}},  {\em Z.Phys.} {\bf C44} (1989) 441.

\bibitem{Weinberg:1980bf}
S.~Weinberg, {\it {Varieties of Baryon and Lepton Nonconservation}},  {\em
  Phys.Rev.} {\bf D22} (1980) 1694.

\bibitem{Davidson:2008bu}
S.~Davidson, E.~Nardi, and Y.~Nir, {\it {Leptogenesis}},  {\em Phys.Rept.} {\bf
  466} (2008) 105--177, [\href{http://xxx.lanl.gov/abs/0802.2962}{{\tt
  arXiv:0802.2962}}].

\bibitem{Hambye:2003rt}
T.~Hambye, Y.~Lin, A.~Notari, M.~Papucci, and A.~Strumia, {\it {Constraints on
  neutrino masses from leptogenesis models}},  {\em Nucl.Phys.} {\bf B695}
  (2004) 169--191, [\href{http://xxx.lanl.gov/abs/hep-ph/0312203}{{\tt
  hep-ph/0312203}}].

\bibitem{Hambye:2005tk}
T.~Hambye, M.~Raidal, and A.~Strumia, {\it {Efficiency and maximal CP-asymmetry
  of scalar triplet leptogenesis}},  {\em Phys.Lett.} {\bf B632} (2006)
  667--674, [\href{http://xxx.lanl.gov/abs/hep-ph/0510008}{{\tt
  hep-ph/0510008}}].

\bibitem{Hambye:2003ka}
T.~Hambye and G.~Senjanovic, {\it {Consequences of triplet seesaw for
  leptogenesis}},  {\em Phys.Lett.} {\bf B582} (2004) 73--81,
  [\href{http://xxx.lanl.gov/abs/hep-ph/0307237}{{\tt hep-ph/0307237}}].

\bibitem{Antusch:2004xy}
S.~Antusch and S.~F. King, {\it {Type II Leptogenesis and the neutrino mass
  scale}},  {\em Phys.Lett.} {\bf B597} (2004) 199--207,
  [\href{http://xxx.lanl.gov/abs/hep-ph/0405093}{{\tt hep-ph/0405093}}].

\bibitem{Froggatt:1978nt}
C.~Froggatt and H.~B. Nielsen, {\it {Hierarchy of Quark Masses, Cabibbo Angles
  and CP Violation}},  {\em Nucl.Phys.} {\bf B147} (1979) 277.

\bibitem{Schwetz:2011zk}
T.~Schwetz, M.~Tortola, and J.~Valle, {\it {Where we are on $\theta_{13}$:
  addendum to 'Global neutrino data and recent reactor fluxes: status of
  three-flavour oscillation parameters'}},  {\em New J.Phys.} {\bf 13} (2011)
  109401, [\href{http://xxx.lanl.gov/abs/1108.1376}{{\tt arXiv:1108.1376}}].

\bibitem{GonzalezGarcia:2010er}
M.~Gonzalez-Garcia, M.~Maltoni, and J.~Salvado, {\it {Updated global fit to
  three neutrino mixing: status of the hints of $\theta_{13}\gtrsim 0$}},  {\em
  JHEP} {\bf 1004} (2010) 056, [\href{http://xxx.lanl.gov/abs/1001.4524}{{\tt
  arXiv:1001.4524}}].

\bibitem{Fogli:2011qn}
G.~Fogli, E.~Lisi, A.~Marrone, A.~Palazzo, and A.~Rotunno, {\it {Evidence of
  $\theta_{13}\gtrsim 0$ from global neutrino data analysis}},  {\em Phys.Rev.}
  {\bf D84} (2011) 053007, [\href{http://xxx.lanl.gov/abs/1106.6028}{{\tt
  arXiv:1106.6028}}].

\bibitem{Altarelli:2010gt}
G.~Altarelli and F.~Feruglio, {\it {Discrete Flavor Symmetries and Models of
  Neutrino Mixing}},  {\em Rev.Mod.Phys.} {\bf 82} (2010) 2701--2729,
  [\href{http://xxx.lanl.gov/abs/1002.0211}{{\tt arXiv:1002.0211}}].

\bibitem{AristizabalSierra:2007ur}
D.~Aristizabal~Sierra, M.~Losada, and E.~Nardi, {\it {Variations on
  leptogenesis}},  {\em Phys.Lett.} {\bf B659} (2008) 328--335,
  [\href{http://xxx.lanl.gov/abs/0705.1489}{{\tt arXiv:0705.1489}}].

\bibitem{AristizabalSierra:2009bh}
D.~Aristizabal~Sierra, L.~A. Munoz, and E.~Nardi, {\it {Purely Flavored
  Leptogenesis}},  {\em Phys.Rev.} {\bf D80} (2009) 016007,
  [\href{http://xxx.lanl.gov/abs/0904.3043}{{\tt arXiv:0904.3043}}].

\bibitem{Sierra:2011vk}
D.~Aristizabal~Sierra and F.~Bazzocchi, {\it {Leptogenesis in the presence of
  exact flavor symmetries}},  {\em JHEP} {\bf 1203} (2012) 057,
  [\href{http://xxx.lanl.gov/abs/1110.3781}{{\tt arXiv:1110.3781}}]. 24 pages,
  7 figures.

\bibitem{Jenkins:2008rb}
E.~E. Jenkins and A.~V. Manohar, {\it {Tribimaximal Mixing, Leptogenesis, and
  $\theta_{13}$}},  {\em Phys.Lett.} {\bf B668} (2008) 210--215,
  [\href{http://xxx.lanl.gov/abs/0807.4176}{{\tt arXiv:0807.4176}}].

\bibitem{Hagedorn:2009jy}
C.~Hagedorn, E.~Molinaro, and S.~Petcov, {\it {Majorana Phases and Leptogenesis
  in See-Saw Models with $A_4$ Symmetry}},  {\em JHEP} {\bf 0909} (2009) 115,
  [\href{http://xxx.lanl.gov/abs/0908.0240}{{\tt arXiv:0908.0240}}].

\bibitem{Bertuzzo:2009im}
E.~Bertuzzo, P.~Di~Bari, F.~Feruglio, and E.~Nardi, {\it {Flavor symmetries,
  leptogenesis and the absolute neutrino mass scale}},  {\em JHEP} {\bf 0911}
  (2009) 036, [\href{http://xxx.lanl.gov/abs/0908.0161}{{\tt
  arXiv:0908.0161}}].

\bibitem{AristizabalSierra:2009ex}
D.~Aristizabal~Sierra, F.~Bazzocchi, I.~de~Medeiros~Varzielas, L.~Merlo, and
  S.~Morisi, {\it {Tri-Bimaximal Lepton Mixing and Leptogenesis}},  {\em
  Nucl.Phys.} {\bf B827} (2010) 34--58,
  [\href{http://xxx.lanl.gov/abs/0908.0907}{{\tt arXiv:0908.0907}}].

\bibitem{Felipe:2009rr}
R.~Felipe and H.~Serodio, {\it {Constraints on leptogenesis from a symmetry
  viewpoint}},  {\em Phys.Rev.} {\bf D81} (2010) 053008,
  [\href{http://xxx.lanl.gov/abs/0908.2947}{{\tt arXiv:0908.2947}}].

\bibitem{Choubey:2010vs}
S.~Choubey, S.~King, and M.~Mitra, {\it {On the Vanishing of the CP Asymmetry
  in Leptogenesis due to Form Dominance}},  {\em Phys.Rev.} {\bf D82} (2010)
  033002, [\href{http://xxx.lanl.gov/abs/1004.3756}{{\tt arXiv:1004.3756}}].

\bibitem{AristizabalSierra:2011ab}
D.~Aristizabal~Sierra, F.~Bazzocchi, and I.~de~Medeiros~Varzielas, {\it
  {Leptogenesis in flavor models with type I and II seesaws}},  {\em
  Nucl.Phys.} {\bf B858} (2012) 196--213,
  [\href{http://xxx.lanl.gov/abs/1112.1843}{{\tt arXiv:1112.1843}}].

\bibitem{Mohapatra:2005ra}
R.~Mohapatra, S.~Nasri, and H.-B. Yu, {\it {Leptogenesis, $\mu-\tau$ symmetry
  and $\theta_{13}$}},  {\em Phys.Lett.} {\bf B615} (2005) 231--239,
  [\href{http://xxx.lanl.gov/abs/hep-ph/0502026}{{\tt hep-ph/0502026}}].

\bibitem{Antusch:2006cw}
S.~Antusch, S.~King, and A.~Riotto, {\it {Flavour-Dependent Leptogenesis with
  Sequential Dominance}},  {\em JCAP} {\bf 0611} (2006) 011,
  [\href{http://xxx.lanl.gov/abs/hep-ph/0609038}{{\tt hep-ph/0609038}}].

\bibitem{Adhikary:2008au}
B.~Adhikary and A.~Ghosal, {\it {Nonzero $U_{e3}$, CP violation and
  leptogenesis in a see-saw type softly broken $A_4$ symmetric model}},  {\em
  Phys.Rev.} {\bf D78} (2008) 073007,
  [\href{http://xxx.lanl.gov/abs/0803.3582}{{\tt arXiv:0803.3582}}].

\bibitem{Lin:2009ic}
Y.~Lin, {\it {A Dynamical approach to link low energy phases with
  leptogenesis}},  {\em Phys.Rev.} {\bf D80} (2009) 076011,
  [\href{http://xxx.lanl.gov/abs/0903.0831}{{\tt arXiv:0903.0831}}].

\bibitem{Branco:2009by}
G.~Branco, R.~Gonzalez~Felipe, M.~Rebelo, and H.~Serodio, {\it {Resonant
  leptogenesis and tribimaximal leptonic mixing with $A_4$ symmetry}},  {\em
  Phys.Rev.} {\bf D79} (2009) 093008,
  [\href{http://xxx.lanl.gov/abs/0904.3076}{{\tt arXiv:0904.3076}}].

\bibitem{Altarelli:2009kr}
G.~Altarelli and D.~Meloni, {\it {A Simplest $A_4$ Model for Tri-Bimaximal
  Neutrino Mixing}},  {\em J.Phys.G} {\bf G36} (2009) 085005,
  [\href{http://xxx.lanl.gov/abs/0905.0620}{{\tt arXiv:0905.0620}}].

\bibitem{Riva:2010jm}
F.~Riva, {\it {Low-Scale Leptogenesis and the Domain Wall Problem in Models
  with Discrete Flavor Symmetries}},  {\em Phys.Lett.} {\bf B690} (2010)
  443--450, [\href{http://xxx.lanl.gov/abs/1004.1177}{{\tt arXiv:1004.1177}}].

\bibitem{Nardi:2007jp}
E.~Nardi, J.~Racker, and E.~Roulet, {\it {CP violation in scatterings, three
  body processes and the Boltzmann equations for leptogenesis}},  {\em JHEP}
  {\bf 0709} (2007) 090, [\href{http://xxx.lanl.gov/abs/0707.0378}{{\tt
  arXiv:0707.0378}}].

\bibitem{Casas:2001sr}
J.~Casas and A.~Ibarra, {\it {Oscillating neutrinos and $\mu \to e \gamma$}},
  {\em Nucl.Phys.} {\bf B618} (2001) 171--204,
  [\href{http://xxx.lanl.gov/abs/hep-ph/0103065}{{\tt hep-ph/0103065}}].

\bibitem{Barbieri:1999ma}
R.~Barbieri, P.~Creminelli, A.~Strumia, and N.~Tetradis, {\it {Baryogenesis
  through leptogenesis}},  {\em Nucl.Phys.} {\bf B575} (2000) 61--77,
  [\href{http://xxx.lanl.gov/abs/hep-ph/9911315}{{\tt hep-ph/9911315}}].

\bibitem{Endoh:2003mz}
T.~Endoh, T.~Morozumi, and Z.-h. Xiong, {\it {Primordial lepton family
  asymmetries in seesaw model}},  {\em Prog.Theor.Phys.} {\bf 111} (2004)
  123--149, [\href{http://xxx.lanl.gov/abs/hep-ph/0308276}{{\tt
  hep-ph/0308276}}]. 26 pages, 8 figures, ptp.style.

\bibitem{Fujihara:2005pv}
T.~Fujihara, S.~Kaneko, S.~K. Kang, D.~Kimura, T.~Morozumi, et~al., {\it
  {Cosmological family asymmetry and CP violation}},  {\em Phys.Rev.} {\bf D72}
  (2005) 016006, [\href{http://xxx.lanl.gov/abs/hep-ph/0505076}{{\tt
  hep-ph/0505076}}].

\bibitem{Nardi:2006fx}
E.~Nardi, Y.~Nir, E.~Roulet, and J.~Racker, {\it {The Importance of flavor in
  leptogenesis}},  {\em JHEP} {\bf 0601} (2006) 164,
  [\href{http://xxx.lanl.gov/abs/hep-ph/0601084}{{\tt hep-ph/0601084}}].

\bibitem{Abada:2006fw}
A.~Abada, S.~Davidson, F.-X. Josse-Michaux, M.~Losada, and A.~Riotto, {\it
  {Flavor issues in leptogenesis}},  {\em JCAP} {\bf 0604} (2006) 004,
  [\href{http://xxx.lanl.gov/abs/hep-ph/0601083}{{\tt hep-ph/0601083}}].

\bibitem{Abada:2006ea}
A.~Abada, S.~Davidson, A.~Ibarra, F.-X. Josse-Michaux, M.~Losada, et~al., {\it
  {Flavour Matters in Leptogenesis}},  {\em JHEP} {\bf 0609} (2006) 010,
  [\href{http://xxx.lanl.gov/abs/hep-ph/0605281}{{\tt hep-ph/0605281}}].

\bibitem{Covi:1996wh}
L.~Covi, E.~Roulet, and F.~Vissani, {\it {CP violating decays in leptogenesis
  scenarios}},  {\em Phys.Lett.} {\bf B384} (1996) 169--174,
  [\href{http://xxx.lanl.gov/abs/hep-ph/9605319}{{\tt hep-ph/9605319}}].

\bibitem{Buchmuller:2004nz}
W.~Buchmuller, P.~Di~Bari, and M.~Plumacher, {\it {Leptogenesis for
  pedestrians}},  {\em Annals Phys.} {\bf 315} (2005) 305--351,
  [\href{http://xxx.lanl.gov/abs/hep-ph/0401240}{{\tt hep-ph/0401240}}].

\bibitem{Blanchet:2006dq}
S.~Blanchet and P.~Di~Bari, {\it {Leptogenesis beyond the limit of hierarchical
  heavy neutrino masses}},  {\em JCAP} {\bf 0606} (2006) 023,
  [\href{http://xxx.lanl.gov/abs/hep-ph/0603107}{{\tt hep-ph/0603107}}].

\bibitem{Ma:1998dx}
E.~Ma and U.~Sarkar, {\it {Neutrino masses and leptogenesis with heavy Higgs
  triplets}},  {\em Phys.Rev.Lett.} {\bf 80} (1998) 5716--5719,
  [\href{http://xxx.lanl.gov/abs/hep-ph/9802445}{{\tt hep-ph/9802445}}].

\bibitem{Strumia:2008cf}
A.~Strumia, {\it {Sommerfeld corrections to type-II and III leptogenesis}},
  {\em Nucl.Phys.} {\bf B809} (2009) 308--317,
  [\href{http://xxx.lanl.gov/abs/0806.1630}{{\tt arXiv:0806.1630}}].

\bibitem{AristizabalSierra:2010mv}
D.~Aristizabal~Sierra, J.~F. Kamenik, and M.~Nemevsek, {\it {Implications of
  Flavor Dynamics for Fermion Triplet Leptogenesis}},  {\em JHEP} {\bf 1010}
  (2010) 036, [\href{http://xxx.lanl.gov/abs/1007.1907}{{\tt
  arXiv:1007.1907}}].

\bibitem{Cerdeno:2006ha}
D.~Cerdeno, A.~Dedes, and T.~Underwood, {\it {The Minimal Phantom Sector of the
  Standard Model: Higgs Phenomenology and Dirac Leptogenesis}},  {\em JHEP}
  {\bf 0609} (2006) 067, [\href{http://xxx.lanl.gov/abs/hep-ph/0607157}{{\tt
  hep-ph/0607157}}].

\bibitem{GonzalezGarcia:2009qd}
M.~Gonzalez-Garcia, J.~Racker, and N.~Rius, {\it {Leptogenesis without
  violation of B-L}},  {\em JHEP} {\bf 0911} (2009) 079,
  [\href{http://xxx.lanl.gov/abs/0909.3518}{{\tt arXiv:0909.3518}}].

\bibitem{Altarelli:2005yx}
G.~Altarelli and F.~Feruglio, {\it {Tri-bimaximal neutrino mixing, $A_4$ and
  the modular symmetry}},  {\em Nucl.Phys.} {\bf B741} (2006) 215--235,
  [\href{http://xxx.lanl.gov/abs/hep-ph/0512103}{{\tt hep-ph/0512103}}].

\bibitem{Low:2003dz}
C.~I. Low and R.~R. Volkas, {\it {Tri-bimaximal mixing, discrete family
  symmetries, and a conjecture connecting the quark and lepton mixing
  matrices}},  {\em Phys.Rev.} {\bf D68} (2003) 033007,
  [\href{http://xxx.lanl.gov/abs/hep-ph/0305243}{{\tt hep-ph/0305243}}].

\bibitem{Chen:2009um}
M.-C. Chen and S.~F. King, {\it {$A_4$ See-Saw Models and Form Dominance}},
  {\em JHEP} {\bf 0906} (2009) 072,
  [\href{http://xxx.lanl.gov/abs/0903.0125}{{\tt arXiv:0903.0125}}].

\bibitem{Varzielas:2010mp}
I.~de~Medeiros~Varzielas and L.~Merlo, {\it {Ultraviolet Completion of Flavour
  Models}},  {\em JHEP} {\bf 1102} (2011) 062,
  [\href{http://xxx.lanl.gov/abs/1011.6662}{{\tt arXiv:1011.6662}}].

\bibitem{Cooper:2011rh}
I.~K. Cooper, S.~F. King, and C.~Luhn, {\it {Renormalisation group improved
  leptogenesis in family symmetry models}},  {\em Nucl.Phys.} {\bf B859} (2012)
  159--176, [\href{http://xxx.lanl.gov/abs/1110.5676}{{\tt arXiv:1110.5676}}].

\bibitem{Bazzocchi:2009qg}
F.~Bazzocchi and I.~de~Medeiros~Varzielas, {\it {Tri-bi-maximal mixing in
  viable family symmetry unified model with extended seesaw}},  {\em Phys.Rev.}
  {\bf D79} (2009) 093001, [\href{http://xxx.lanl.gov/abs/0902.3250}{{\tt
  arXiv:0902.3250}}].

\bibitem{deMedeirosVarzielas:2011tp}
I.~de~Medeiros~Varzielas, R.~Gonzalez~Felipe, and H.~Serodio, {\it {Leptonic
  mixing, family symmetries and neutrino phenomenology}},  {\em Phys.Rev.} {\bf
  D83} (2011) 033007, [\href{http://xxx.lanl.gov/abs/1101.0602}{{\tt
  arXiv:1101.0602}}].

\bibitem{AristizabalSierra:2009mq}
D.~Aristizabal~Sierra, M.~Losada, and E.~Nardi, {\it {Lepton Flavor
  Equilibration and Leptogenesis}},  {\em JCAP} {\bf 0912} (2009) 015,
  [\href{http://xxx.lanl.gov/abs/0905.0662}{{\tt arXiv:0905.0662}}].

\end{thebibliography}\endgroup
\end{document}